\RequirePackage{fix-cm}
\documentclass[smallextended]{svjour3}       

\smartqed  % flush right qed marks, e.g. at end of proof

\usepackage[hyphens]{url}  % Allows line breaks in URLs

\usepackage{array} % adiciona \newcolumntype e \arraybackslash
\newcolumntype{L}[1]{>{\raggedright\arraybackslash}p{#1}}
\usepackage{colortbl}
\usepackage{graphicx}         % para \resizebox
\usepackage[table]{xcolor}    % para \rowcolor e \rowcolors
\usepackage{mdframed}         % para mdframed
\usepackage{pgfplots}
\usepgfplotslibrary{groupplots}
\pgfplotsset{compat=1.18}
\usetikzlibrary{positioning}
\usepackage[dvipsnames]{xcolor}
\usepackage{tikz}
\usetikzlibrary{calc}
\pgfplotsset{compat=1.17}
\usepackage{listings}
\usepackage{tabularx}
\usepackage[most]{tcolorbox}
\definecolor{myblue}{RGB}{8,81,156}
\usepackage{pgfplots}
\pgfplotsset{compat=1.18}
\usepgfplotslibrary{statistics,groupplots}

\usepackage{booktabs}
\usepackage{array}
\newcolumntype{Y}{>{\raggedright\arraybackslash}X}
\newtcolorbox{modeloutput}{
  enhanced,
  boxrule=0.4pt,
  arc=2pt,
  left=6pt,
  right=6pt,
  top=4pt,
  bottom=4pt,
  colback=cyan!4,        % fundo bem claro
  colframe=cyan!35,      % borda suave
  fontupper=\itshape
}

\usepackage{lineno}
\usepackage[colorlinks=true]{hyperref}
\modulolinenumbers[5]

\usepackage{booktabs}   %% For formal tables:
\usepackage{subcaption} %% For complex figures with subfigures/subcaptions
\usepackage[labelfont=bf]{caption}
\usepackage[utf8]{inputenc} 
\usepackage[T1]{fontenc}
\usepackage{microtype}

\usepackage{enumitem}
\usepackage{color}
\usepackage{amssymb}
\usepackage{multirow}
\usepackage{etoolbox}
\usepackage[plain]{algorithm}% http://ctan.org/pkg/algorithm
\usepackage[noend]{algorithmic}% http://ctan.org/pkg/algorithm
\usepackage{syntax}
\usepackage{comment}
\usepackage{amsmath}
\usepackage{xspace}
\usepackage{tcolorbox}
\usepackage[all]{nowidow}
\usepackage{booktabs}
\usepackage{algorithmic}
\usepackage[english]{babel}
\usepackage{hyphenat}
\usepackage{ulem}
\usepackage{listings}
\usepackage{mdframed}

\definecolor{pblue}{rgb}{0.13,0.13,1}
\definecolor{pgreen}{rgb}{0,0.5,0}
\definecolor{pred}{rgb}{0.9,0,0}
\definecolor{pgrey}{rgb}{0.46,0.45,0.48}
\definecolor{darkblue}{rgb}{0.0, 0.0, 0.55}
\definecolor{light-gray}{gray}{0.9}

\newcommand{\lstbg}[3][0pt]{{\fboxsep#1\colorbox{#2}{\strut #3}}}

\lstdefinelanguage{diff}{
  basicstyle=\ttfamily\small,
  morecomment=[f][\lstbg{pred!20}]-,
  morecomment=[f][\lstbg{pgreen!20}]+,
  morecomment=[f][\textit]{@@},
  morecomment=[f][\textit]{---},
  morecomment=[f][\textit]{+++},
}

\hypersetup{
    colorlinks=true,
    linkcolor=blue,
    filecolor=magenta,      
    urlcolor=blue,
    citecolor=violet,
    pdfpagemode=FullScreen,
}

\newcommand{\gptoss}{\textsc{GPT-OSS-20B}}
\newcommand{\gptfive}{\textsc{GPT-5.4}}
\newcommand{\gemini}{\textsc{Gemini-3.1-Pro-Preview}}
\newcommand{\gptfiveAcc}{93.8\%}
\newcommand{\gpttossAcc}{80.5\%}
\newcommand{\claudeAcc}{94.7\%}
\newcommand{\geminiAcc}{99.6\%}
\newcommand{\refactTypes}{47}

\newcommand{\qwen}{\textsc{Qwen-3.6-35B}}
\newcommand{\phimodel}{\textsc{Phi-4-14B}}
\newcommand{\llama}{\textsc{Llama-3.2-3B}}
\newcommand{\claude}{\textsc{Claude-4.6-Sonnet}}
\newcommand{\numTransformationsReal}{44}
\newcommand{\openjdk}{\textsc{OpenJDK-Temurin-21.0.7+6}}

\newcommand{\bugs}{226} %Java
\newcommand{\bugsCE}{185} %Java
\newcommand{\bugsBC}{41} %Java

\newcommand{\gemmaNew}{\textsc{Gemma-4-31B}}

\newcommand{\revision}[1]{{\color{black}#1\normalfont}}
\newcommand{\revisionTwo}[1]{{\color{black}#1\normalfont}}
\definecolor{green}{RGB}{0,100,0}

\newcommand{\eclipse}{\textsc{Eclipse}}

\newcommand{\netbeans}{\textsc{NetBeans}}
\newcommand{\intellij}{\textsc{IntelliJ-IDEA}}
\newcommand{\vscode}{\textsc{VSCode}}

\newcommand{\randoop}{\textsc{Randoop}}
\newcommand{\saferefactor}{\textsc{SafeRefactor}}
\newcommand{\jdolly}{\textsc{JDolly}}

\usepackage{subcaption}
\usepackage{xcolor}
\usepackage{listings}
\lstdefinestyle{refjava}{
  language=Java,
  basicstyle=\scriptsize\ttfamily,      % compacta sem perder legibilidade
  numberstyle=\tiny,
  numbersep=6pt,
  frame=single,
  rulecolor=\color{black!30},
  xleftmargin=1.25em,
  framexleftmargin=1.0em,
  aboveskip=4pt,
  belowskip=2pt,
  columns=fullflexible,
  keepspaces=true,
  tabsize=2,
  showstringspaces=false,
  breaklines=true,
  captionpos=b
}

\journalname{}

\begin{document}

\title{
Foundation Models as Oracles for Refactoring Correctness Detection
}

\author{Rohit Gheyi \and Rian Melo \and Jonhnanthan Oliveira \and M\'{a}rcio Ribeiro \and Baldoino Fonseca
}

\institute{R. Gheyi and R. Melo and J. Oliveira\at
              Federal University of Campina Grande \\
              \email{rohit@dsc.ufcg.edu.br, rian.melo@ccc.ufcg.edu.br, jonhnanthan@copin.ufcg.edu.br}             \\
           \and
           M. Ribeiro and B. Fonseca\at
           Federal University of Alagoas \\
           \email{marcio@ic.ufal.br, baldoino@ic.ufal.br}
}  

\date{}

\maketitle

\begin{abstract}
Refactoring tools in popular Integrated Development Environments (IDEs) can introduce unintended behavioral changes or compilation errors, a persistent challenge that undermines developer trust in automated transformations. Traditional detection approaches rely on handcrafted preconditions and static and dynamic analyses, yet remain limited in adaptability and can miss subtle \revision{correctness issues}.
\revision{
This study examines the potential of foundation models to serve as oracles for detecting refactoring bugs in Java programs. We evaluate zero-shot prompting, without task-specific training, across \bugs{} real refactoring bugs collected over more than a decade from widely used Java IDEs (\intellij{}, \eclipse{}, and \netbeans{}), spanning \refactTypes{} refactoring types.
}
\revision{
Our results indicate that foundation models can be effective for this task, although performance varies across models. In the first-run setting, \gptoss{} achieved \gpttossAcc{} accuracy, while \gptfive{} reached \gptfiveAcc{}. We also evaluated other \revisionTwo{open-weight} and proprietary models: \gemmaNew{} achieved the strongest result among \revisionTwo{open-weight} models, and \gemini{} achieved the best overall result among all evaluated models. The two primary models were evaluated over five attempts, whereas the additional models were evaluated in a single-run comparison. \revisionTwo{Metamorphic testing indicates that model predictions remain largely consistent under the tested semantics-preserving perturbations, but these results should be interpreted as robustness evidence rather than as evidence against memorization or data contamination.}
}
Beyond detection accuracy, foundation models can provide short explanations that may help support developer inspection, \revision{operate} across refactoring types without \revision{explicitly encoded refactoring-specific} rules, and \revision{may serve} as lightweight triage \revision{aids} in development workflows. \revision{Our findings suggest that foundation models can complement traditional refactoring checks by flagging suspicious transformations for developer inspection.}
\end{abstract}


\section{Introduction}

Refactoring~\cite{Fowler-book-1999,Opdyke-PHD-1992}, the systematic restructuring of code to improve its internal quality while preserving observable behavior, has become a cornerstone of modern software development. Since Roberts~\cite{Roberts-PHD-1999} introduced the first automated refactoring tool, popular Integrated Development Environments (IDEs) such as \intellij{}, \eclipse{}, and \netbeans{} have integrated refactoring implementations that make complex transformations accessible to \revision{large developer communities}. This automation is not merely a convenience \revision{but an important form of support}: recent empirical studies reveal that over 40\% of developers perform refactorings daily~\cite{golubev-fse-2021}, making automated support \revision{important} for maintaining code quality and \revision{supporting} the long-term sustainability of increasingly complex software systems.

Despite decades of research and engineering effort, automated refactoring tools \revision{remain imperfect}: they \revision{can still} introduce subtle behavioral changes that violate the core promise of behavior preservation. This challenge represents more than a technical inconvenience\revision{:} it undermines developer trust and can lead to production failures that escape both code review and existing test suites. The difficulty stems from the inherent complexity of ensuring \revision{behavior preservation across} program transformations, where seemingly straightforward changes can have far-reaching and unexpected consequences.

Traditional approaches to refactoring correctness rely on handcrafted preconditions and extensive static and dynamic analyses~\cite{Soares-TSE-2013,test-tools-fse07,steimann-ecoop}. Tools like \saferefactor{}~\cite{saferefactor-ieee} generate and execute test cases to expose unintended changes, while others implement complex, transformation-specific analyses~\cite{Schafer-OOPSLA-2010}. However, these approaches face \revision{important} limitations: \revision{defining a sound and complete set of preconditions is challenging~\cite{Schafer-PLPV-2009}, implementing the corresponding analyses requires substantial engineering effort, and proposing new analyses for additional refactoring scenarios is far from straightforward. Moreover, such approaches are difficult to adapt to new refactoring types, must continuously evolve alongside programming languages, or both.} 

The complexity is compounded by the rich semantics of modern programming languages. While individual language constructs may appear straightforward in isolation, their interactions create a combinatorial explosion of edge cases. Java exemplifies this challenge with its numerous traps, pitfalls, and corner cases~\cite{bloch-java-puzzlers} involving inheritance, method resolution, exception handling, and type inference. This complexity may explain why many developers continue to prefer manual refactorings despite the availability of automated tools~\cite{Tempero-ACM-2017}. They simply do not trust automated refactoring implementations to preserve program behavior reliably.

The emergence of foundation models has catalyzed a paradigm shift in software engineering, demonstrating \revision{strong performance} in code understanding, generation, and analysis tasks~\cite{se-llms-2023,wang2024}. Unlike traditional rule-based approaches that require explicit specification of every possible edge case, foundation models \revision{can exhibit capabilities that may support reasoning-like behavior and} allow them to \revision{perform across} diverse scenarios without task-specific training. This generalization potential is particularly compelling for refactoring validation, where the space of possible transformations and their interactions is vast and evolving.
\revision{
In this article, we use the term foundation models to refer to general-purpose models trained on broad data and adaptable to many downstream tasks. Large language models (LLMs) are a subclass of foundation models specialized in processing and generating natural language.
}
They can analyze code at multiple levels of abstraction simultaneously, \revision{identify semantic relationships that may not be explicitly encoded in traditional analyses}, and provide natural language \revision{rationales} for their decisions, helping developers \revision{inspect} why a transformation might be problematic rather than simply flagging it as incorrect. \revision{In addition, some \revisionTwo{open-weight} models can be executed locally or on low-cost cloud hardware, making them attractive candidates for integration into development workflows where cost, latency, and privacy matter.}
However, the application of foundation models to refactoring correctness remains \revision{relatively} unexplored. While these models have shown promise in various software engineering tasks~\cite{se-llms-2023}, their effectiveness in detecting the subtle behavioral changes that characterize refactoring bugs, particularly their ability to distinguish between surface-level code changes and \revision{potential behavior-changing differences}, represents an open research question~\cite{DBLP:conf/fose-ws/FanGHLSYZ23}.

This article investigates the extent to which foundation models \revision{may serve as complementary oracle-like components} for detecting behavioral changes \revision{(BCs) and compilation errors (CEs)} introduced by Java program refactorings. We address this question through \revision{an} empirical study that evaluates foundation models on real-world refactoring bugs rather than synthetic examples, providing \revision{evidence about their possible use in development settings}.
Our evaluation encompasses \bugs{} refactoring bugs collected over more than a decade from three widely used Java IDEs (\intellij{}, \eclipse{}, \netbeans{}), spanning \refactTypes{} distinct refactoring types. This dataset represents \revision{a broad sample of} refactoring challenges encountered in practice, including both compilation errors and subtle behavioral changes that traditional tools often miss. We employ zero-shot prompting \revision{in the main benchmark, and use MetaPrompting to derive a complementary diff-oriented prompt for large-project feasibility analysis. This allows us to assess model behavior without domain-specific training while also exploring how the approach behaves when full project context is unavailable.}

\revision{
Our key findings indicate that foundation models can achieve encouraging accuracy in refactoring-bug detection. In the first-run setting, the \revisionTwo{open-weight} model \gptoss{} attains \gpttossAcc{} accuracy, while the proprietary \gptfive{} model reaches \gptfiveAcc{}. Additional proprietary and \revisionTwo{open-weight} models show that performance varies across model families. \revisionTwo{
Metamorphic testing shows that predictions remain largely stable under the tested behavior-preserving perturbations, but we interpret this result as robustness evidence rather than as evidence against memorization or data contamination.
}
}
\revision{A possible cost-conscious deployment strategy is suggested by the results: use inexpensive local or \revisionTwo{open-weight} models for initial screening and selectively employ stronger proprietary models for uncertain or high-risk cases, balancing accuracy, latency, and resource utilization.}

The implications of our findings may extend beyond refactoring to software engineering tasks that require \revision{reasoning about} \revision{behavioral preservation}. Compiler-optimization validation, mutation testing, and program-transformation verification face the same core challenge of distinguishing behavior-preserving from behavior-changing transformations~\cite{DBLP:conf/icse/SteimannT10,demillo1978hints,jia2011analysis,aho1986compilers}. Our results suggest that foundation models \revision{may serve as complementary oracle-like components} for semantic equivalence \revision{assessment} across these domains.
The integration of foundation models into modern development environments, exemplified by AI-enabled IDEs such as Antigravity~\cite{antigravity}, Windsurf~\cite{windsurf} and Cursor~\cite{cursor}, offers \revision{a plausible path for further integration}. Rather than replacing existing verification techniques, foundation models can augment them by providing \revision{explanation-oriented} triage that \revision{may reduce} the computational burden on heavyweight analyses while improving developers' understanding of transformation risks.
\revision{Open challenges remain, including context-window limits, cost and latency, model non-determinism, and the choice of behavioral-equivalence criteria. Nevertheless, the results indicate that foundation models are a promising complement to existing refactoring-validation techniques. All artifacts are publicly available online.\footnote{\url{https://zenodo.org/records/21910261}}
}

This article is organized as follows. Section~\ref{sec:example} presents a motivating example that illustrates the challenges of refactoring correctness. Section~\ref{sec:methodology} details our experimental methodology, including dataset construction, model selection, and evaluation metrics. Section~\ref{sec:results} presents our empirical findings across different models and refactoring types. Section~\ref{sec:discussion} discusses our results. Section~\ref{sec:relwork} situates our work within the broader landscape of refactoring research and foundation model applications. Finally, Section~\ref{sec:conclusion} concludes with a synthesis of our contributions and directions for future research.

\section{Motivating Example}
\label{sec:example}

Ensuring that a refactoring transformation preserves program correctness requires satisfying a set of \textit{preconditions}. These preconditions \revision{are intended to ensure} that the resulting program compiles and preserves the original behavior. For instance, in the \textit{Extract Class} refactoring~\cite{Fowler-book-1999}, the newly introduced class must not conflict with an existing class name. In practice, developers may perform refactorings manually, an error-prone and time-consuming process, or rely on automated tools available in IDEs such as \vscode{}~\cite{vscode}, \intellij{}~\cite{intellij}, \eclipse{}~\cite{eclipse}, and \netbeans{}~\cite{netbeans}. 

Behavior preservation is typically verified by compiling the refactored program and executing its test suite~\cite{Fowler-book-1999}. While this strategy is \revision{often} effective at identifying compilation errors, it \revision{may fail} to capture more subtle behavioral deviations. Rachatasumrit and Kim~\cite{DBLP:conf/icsm/RachatasumritK12} find that, in practice, many developers’ test suites are insufficient, as they do not adequately exercise \revision{many} refactored methods and fields. Moreover, \revision{many} refactoring implementations check only a subset of the necessary preconditions rather than encoding them comprehensively, which increases the likelihood of correctness violations.

\begin{figure*}[htbp]
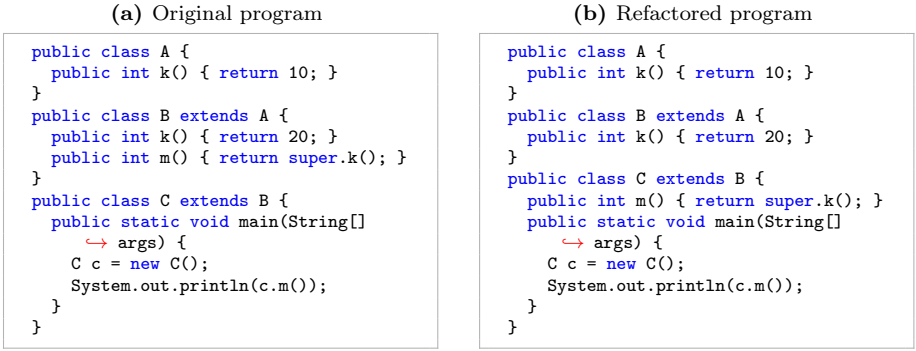

  \centering
  \begin{subfigure}[t]{0.47\textwidth}
    \caption{Original program}
    \label{lst:orig2}
\begin{lstlisting}[style=refjava]
public class A {
  public int k() { return 10; }
}
public class B extends A {
  public int k() { return 20; }
  public int m() { return super.k(); }
}
public class C extends B {
  public static void main(String[] args) {
    C c = new C();
    System.out.println(c.m());
  }
}
\end{lstlisting}
  \end{subfigure}
  \hfill
  \begin{subfigure}[t]{0.47\textwidth}
    \caption{Refactored program}
    \label{lst:refac20}
\begin{lstlisting}[style=refjava]
public class A {
  public int k() { return 10; }
}
public class B extends A {
  public int k() { return 20; }
}
public class C extends B {
  public int m() { return super.k(); }
  public static void main(String[] args) {
    C c = new C();
    System.out.println(c.m());
  }
}
\end{lstlisting}
  \end{subfigure}
  \caption{Applying \textit{Push Down Method} \texttt{B.m} to \texttt{C} using \eclipse{} JDT introduces a behavioral change.}
  \label{fig:exampleOWC}
\end{figure*}

To illustrate, consider the program in Figure~\ref{fig:exampleOWC}(a), where method \texttt{m} in class \texttt{B} invokes \texttt{super.k()}. Applying the \textit{Push Down Method} refactoring in \eclipse{} JDT moves \texttt{B.m} to subclass \texttt{C}, producing the program in Figure~\ref{fig:exampleOWC}(b). Although compilation succeeds, the behavior changes: executing \texttt{C.main} prints \texttt{10} before refactoring and \texttt{20} afterwards. This \eclipse{} bug\footnote{\url{https://bugs.eclipse.org/bugs/show_bug.cgi?id=356698}} occurs because the refactoring fails to update the call to \texttt{super.k()}, causing the call to resolve to a different method implementation.

\begin{figure*}[htbp]
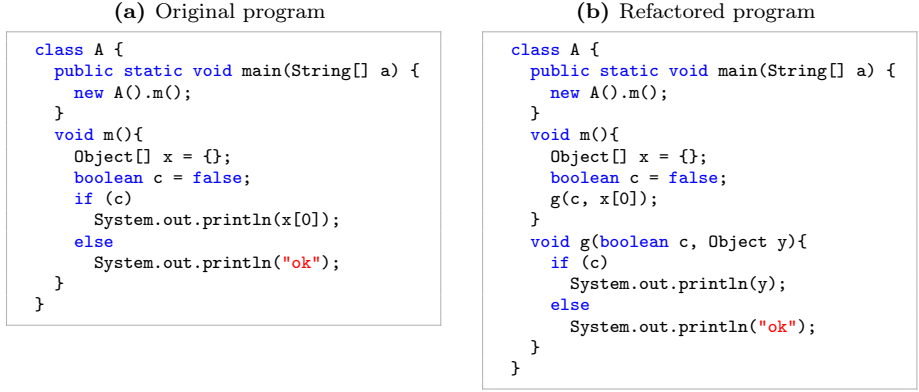

  \centering
  \begin{subfigure}[t]{0.47\textwidth}
    \caption{Original program}
    \label{lst:orig2-novo}
\begin{lstlisting}[style=refjava]
class A {
  public static void main(String[] a) { 
    new A().m(); 
  }
  void m(){
    Object[] x = {}; 
    boolean c = false;
    if (c) 
      System.out.println(x[0]); 
    else 
      System.out.println("ok");
  }
}
\end{lstlisting}
  \end{subfigure}
  \hfill
  \begin{subfigure}[t]{0.47\textwidth}
    \caption{Refactored program}
    \label{lst:refac21}
\begin{lstlisting}[style=refjava]
class A {
  public static void main(String[] a) { 
    new A().m(); 
  }
  void m(){
    Object[] x = {}; 
    boolean c = false;
    g(c, x[0]); 
  }
  void g(boolean c, Object y){
    if (c) 
      System.out.println(y); 
    else 
      System.out.println("ok");
  }
}
\end{lstlisting}
  \end{subfigure}
  \caption{Applying \textit{Extract Method} using \intellij{} introduces a behavioral change.}
  \label{fig:exampleOWC11}
\end{figure*}

As another example, consider applying \textit{Extract Method} in \intellij{} to the conditional inside \texttt{m}. In the original program (Figure~\ref{fig:exampleOWC11}(a)), the execution prints \texttt{ok} because the array access \texttt{x[0]} is protected by the guard condition \texttt{c}. After refactoring (Figure~\ref{fig:exampleOWC11}(b)), however, \intellij{} with Fold parameters enabled extracts the expression \texttt{x[0]} as an argument in the call \texttt{g(c, x[0])}. This transformation alters the original control flow: since Java eagerly evaluates method arguments, \texttt{x[0]} is now evaluated unconditionally before entering the method, leading to an \texttt{ArrayIndexOutOfBoundsException} that did not occur in the original code. This behavioral change constitutes a correctness bug in the refactoring tool, which was subsequently reported to \intellij{}.\footnote{\url{https://youtrack.jetbrains.com/issue/IDEA-92815/Extract-method-refactoring-with-Fold-parameters-is-too-clever-breaks-functioning-code}}
While this issue may be obvious in such a small program, detecting similar problems in larger, real-world codebases is \revision{more challenging}. \revision{Assessing} the correctness of a transformation requires careful handling of subtle language semantics and numerous corner cases, which is a nontrivial task. Even experienced \intellij{} developers were unable to understand the behavioral change described in the initial bug report.

Although simple, this example mirrors issues reported in real-world systems. Gligoric et al.~\cite{Gligoric-ecoop-13} studied five large open-source Java projects and uncovered 77 refactoring-related bugs in \eclipse{}, many resembling the problem in Figure~\ref{fig:exampleOWC}. Such bugs are not isolated: IDEs including \intellij{}, \vscode{}, and \netbeans{} have exhibited similar issues due to the inherent difficulty of defining complete correctness preconditions. Formalizing refactoring semantics is \revision{difficult}~\cite{Schafer-PLPV-2009}, leading many tools to implement only partial checks. As a result, developers may distrust automated tools and prefer manual refactorings~\cite{Tempero-ACM-2017}.

Prior research has proposed several techniques to test and improve the correctness of refactoring implementations~\cite{Soares-TSE-2013,test-tools-fse07,steimann-ecoop,mongiovi-scp-2014,dong-icse-2025}. For example, \saferefactor{}~\cite{saferefactor-ieee} automatically generates test cases to detect unintended behavioral changes introduced by transformations, while other approaches rely on \revision{specialized} type-specific analyses and carefully designed refactoring conditions~\cite{Schafer-OOPSLA-2010}. Although effective in some settings, these techniques are often labor-intensive and costly to adapt to new refactoring types. \revision{Moreover, their effectiveness depends on the adequacy~\cite{DBLP:journals/tse/GoodenoughG75} of the underlying validation mechanisms: existing test suites are often insufficient to reveal all behavioral changes introduced by refactorings~\cite{DBLP:conf/icsm/RachatasumritK12}, and static or dynamic analyses may fail to capture subtle interactions among language constructs.}

Refactoring implementations must \revision{account for} a \revision{broad} range of programming-language constructs, which is far from trivial. Even when the semantics of individual constructs seem simple in isolation, their interactions frequently give rise to subtle and unexpected behaviors. Java, for instance, is well known for its many traps, pitfalls, and corner cases, which make \revision{assessing} program transformations particularly challenging~\cite{bloch-java-puzzlers}. This complexity makes it difficult to define and enforce complete preconditions for all refactorings. As a result, \revision{existing tools do not reliably detect all} behavior-changing transformations, and refactoring correctness remains a persistent open problem.

More recently, foundation models have emerged as \revision{promising complementary techniques} for software analysis. \revision{Rather than replacing existing validation mechanisms used by developers and refactoring tools, such as compiler checks, static analyses, and regression testing, we position foundation models as an additional oracle-like check over the transformation.} In contrast to traditional static and dynamic techniques, which are often computationally expensive and tightly coupled to language-specific semantics, foundation models \revision{may provide relatively lightweight analyses, in some cases} on consumer-grade hardware~\cite{se-llms-2023,wang2024}. \revision{Such analyses can provide additional evidence about whether a transformation may introduce a compilation error or a behavioral change, especially when available tests or handcrafted analyses are incomplete.} However, their effectiveness in detecting refactoring-related bugs, especially those involving subtle behavioral changes, remains, to the best of our knowledge, an open research question~\cite{DBLP:conf/fose-ws/FanGHLSYZ23}.

\section{Research Method}
\label{sec:methodology}

\begin{figure*}[t]
    \centering
    \includegraphics[width=\textwidth]{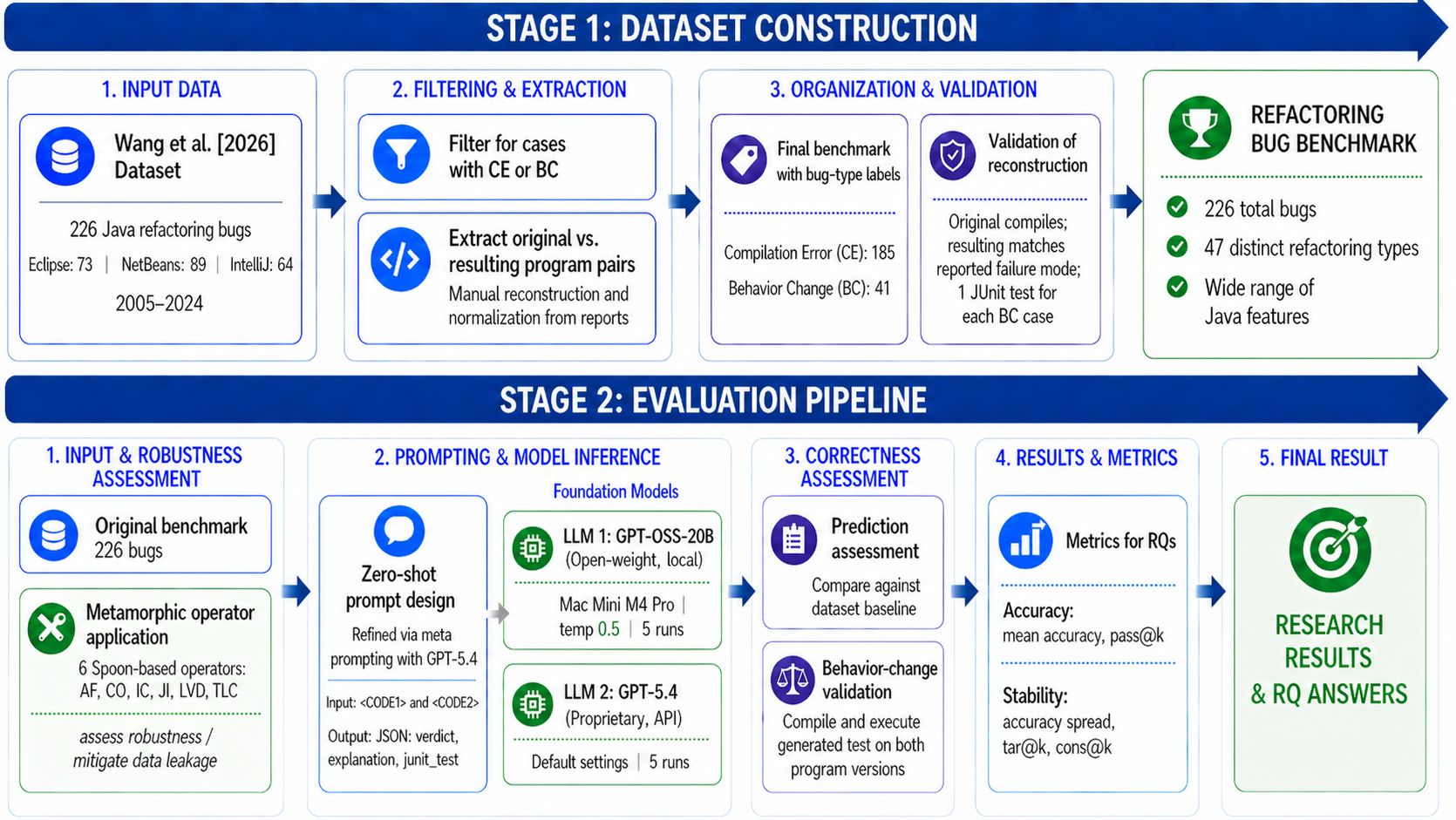}
    \caption{Overview of the research method. Stage~1 summarizes dataset construction, including filtering, reconstruction, and validation of \bugs{} refactoring bugs spanning multiple refactoring types. Stage~2 summarizes the evaluation pipeline, including stability and sensitivity assessment with metamorphic operators, prompt-based model inference with \gptoss{} and \gptfive{}, correctness assessment, and computation of accuracy and stability metrics.}
    \label{fig:method-overview}
\end{figure*}

This section details the research method employed in our study.
\revision{
Figure~\ref{fig:method-overview} presents an overview of the research method, which is detailed next.}

\subsection{Research Goal and Questions}

The objective of this study is defined as follows: analyze the accuracy of foundation models for the purpose of detecting refactoring bugs, including compilation errors and behavioral changes, with respect to previously reported refactoring-related bugs from the point of view of software engineering researchers. We define the following research questions:

\begin{description}
    \item[RQ$_{1}$] To what extent can an \revisionTwo{open-weight} foundation model (\gptoss{}) detect refactoring bugs, including compilation errors and behavioral changes?
    \item[RQ$_{2}$] To what extent can a proprietary foundation model (\gptfive{}) detect refactoring bugs, including compilation errors and behavioral changes?
    \item[RQ$_{3}$] To what extent does the accuracy of these models remain stable under the semantics-preserving metamorphic transformations evaluated in this study?  
\end{description}

\subsection{Dataset}
\label{sec:dataset}

We evaluate \bugs{} refactoring bugs reported across major Java IDEs, namely \intellij{} (Mar~2016--Mar~2024), \eclipse{} (Mar~2005--Jun~2023), and \netbeans{} (Apr~2012--Apr~2024). Our study builds on the dataset organized by Wang et al.~\cite{wang2024empiricalstudyrefactoringengine}, from which we retain cases where the transformation introduced compilation errors \revision{(CEs)} or behavioral changes \revision{(BCs)}. We extend the dataset of Wang et al.~\cite{wang2024empiricalstudyrefactoringengine}, a minor contribution of this article, by extracting, for each report, both the original Java program before the refactoring and the resulting Java program produced after the faulty transformation.

\revision{
For several reports, this process required manual reconstruction because the original submissions were not directly executable artifacts, but rather bug reports mixing code fragments, natural-language explanations, and follow-up discussion~\cite{good-bug-report-tse-2010}. In some cases\footnote{ID 97: \url{https://github.com/eclipse-jdt/eclipse.jdt.ui/issues/1530}}, the bug report already contained a complete executable example, which we reused in our dataset after minor normalization steps, such as removing comments. In other cases\footnote{ID 257: \url{https://bugs.eclipse.org/bugs/show_bug.cgi?id=87483}}, the report provided only partial code snippets embedded in the description or discussion thread; for those instances, we reconstructed the missing class declarations and surrounding program context from the reported fragments and the clarifications provided in follow-up comments.
}

\revision{
Our final dataset contains \bugs{} bugs in total, including \bugsCE{} compilation errors and \bugsBC{} behavioral changes. Regarding tool distribution, \eclipse{} accounts for 73 bugs, \netbeans{} for 89, and \intellij{} for 64. The original program size of the examples ranges from 3 to 28 LOC, with an average of 10.8 LOC per instance. To minimize bias in model evaluation, we removed all code comments from the examples. We also performed a sanity check to avoid redundancy by collapsing similar bug reports that affect multiple tools into a single representative case.
}

\revision{
For every instance in the dataset, we store both Java versions involved in the refactoring scenario: the original program and the resulting program. To validate the reconstruction, we compiled all original and resulting programs using \texttt{javac} from \openjdk{} and stored the corresponding compiler logs in the dataset. Every original program compiles successfully. For instances labeled as compilation errors, the resulting program fails to compile, consistently with the reported bug; the corresponding compiler diagnostics provide executable evidence for the compilation-error label. 
}

\revision{
For instances labeled as behavior changes, the resulting program compiles successfully, again consistently with the report, but its behavior differs from that of the original program. In this case, we additionally provide one JUnit test class containing a single test case that exposes the behavioral difference. Each such test case was compiled successfully against both the original and the resulting programs using the same Java environment, and we store the corresponding test-compilation logs. We then executed the same test unchanged on both versions and stored the execution results. In all \bugsBC{} cases, the test passes on the original program and fails on the resulting program, thereby confirming the reported behavior change after refactoring through executable evidence. None of these test cases relies on Java reflection.
}

\revision{
The Java programs in our dataset exercise a broad range of Java features that commonly interact with refactoring behavior. They include fundamental object-oriented constructs such as classes, packages, imports, access modifiers, fields, methods, constructors, inheritance, interfaces, abstract classes, method overloading and overriding, field hiding, and qualified member access through \texttt{this}, \texttt{super}, and enclosing instances. The programs also cover several forms of nested and local declarations, including member classes, static nested classes, local classes, anonymous classes, nested interfaces, and enums. The dataset further includes common control-flow and expression-level constructs, such as conditionals, loops, enhanced \texttt{for} loops, \texttt{switch} statements and switch expressions, ternary expressions, array accesses, varargs, casts, \texttt{instanceof} checks, assignments with side effects, and try-catch and try-with-resources statements. It also exercises type-system features such as generics, bounded type parameters, raw types, wildcards, generic methods, arrays, primitive and boxed types, and final fields and variables. In addition, several programs rely on annotations, static imports, static members, initializer blocks, synchronization, exceptions, and standard-library APIs. The programs also include more recent Java constructs, including lambda expressions, method references, functional interfaces, default and static interface methods, \texttt{var}-based local type inference, and switch expressions with \texttt{yield}.
}

\revision{
The dataset also spans a broad and diverse spectrum of refactorings reported by practitioners. Table~\ref{tab:top-refactorings} summarizes the ten most frequent refactoring types in the dataset, with all remaining cases grouped into \textit{Others}. The most frequent categories are \textit{Move Method}, \textit{Inline Method}, \textit{Pull Up Method}, \textit{Extract Local Variable}, \textit{Rename Method}, \textit{Change Method Signature}, \textit{Extract Method}, \textit{Move Class}, \textit{Inline Variable}, and \textit{Push Down Method}. Beyond these common categories, the dataset also includes \textit{Add Parameter}, \textit{Encapsulate Field}, \textit{Pull Up Field}, \textit{Push Down Field}, \textit{Rename Class}, \textit{Rename Field}, \textit{Rename Variable}, \textit{Introduce Variable}, \textit{Pull Up Interface}, \textit{Introduce Constant}, \textit{Extract Superclass}, \textit{Generalize Type}, \textit{Join Variable Declaration}, \textit{Extract Variable}, \textit{Introduce Factory}, \textit{Extract Class}, \textit{Inline Field}, \textit{Convert Anonymous to Nested Class}, \textit{Refactoring to Conditional}, \textit{Introduce Field}, \textit{Extract Variable as Enum}, \textit{Replace Inheritance with Delegation}, \textit{Replace Anonymous with Lambda}, \textit{Convert to AtomicBoolean}, \textit{Delete Unused Variable}, \textit{Extract Parameter}, \textit{Extract Interface}, \textit{Replace Constructor with Factory}, \textit{Convert Boolean}, \textit{Delete Unused Method}, \textit{Convert to Instance Method}, \textit{Type Migration}, \textit{Inline Interface}, \textit{Change Class Signature}, \textit{Move Field}, \textit{Move Inner to Outer Level}, and \textit{Replace Constructor with Builder}. This diversity is important because it exposes the models to both frequent and less common transformations, ranging from method- and field-level changes to broader type- and design-level restructurings.

}

\begin{table}[t]
\centering
\caption{Top 10 refactoring types in the dataset.}
\label{tab:top-refactorings}
\small
\begin{tabular}{lr}
\hline
\textbf{Refactoring} & \textbf{Instances} \\
\hline
\textit{Move Method} & 32 \\
\textit{Inline Method} & 20 \\
\textit{Pull Up Method} & 18 \\
\textit{Extract Local Variable} & 15 \\
\textit{Rename Method} & 14 \\
\textit{Change Method Signature} & 13 \\
\textit{Extract Method} & 12 \\
\textit{Move Class} & 11 \\
\textit{Inline Variable} & 10 \\
\textit{Push Down Method} & 6 \\
\textit{Others} & 75 \\
\hline
\end{tabular}
\end{table}

\subsection{Behavior Preservation Notion}
\label{sec:behavior-preservation}

\revision{
In this study, we adopt a client-observable notion of behavioral equivalence tailored to the analysis of refactoring bugs. This notion is aligned with the equivalence notion used in \saferefactor{}~\cite{saferefactor-ieee}. Given an initial program and a resulting program produced after a refactoring, behavior is considered to be preserved only if the resulting program remains compilable and both versions exhibit the same observable behavior for the exercised public APIs and externally visible effects of the methods they have in common.
Thus, if the initial program compiles but the resulting program does not, we treat the transformation as behavior-changing, since client code can no longer execute the transformed program. For pairs in which both versions compile, two programs are considered behaviorally equivalent if the same client-level code, executed unchanged against both versions, produces the same return values, printed output, thrown exceptions, and externally visible state changes when exercising their public interface. This definition deliberately focuses on effects visible to clients of the program, rather than on internal implementation details, since refactorings are expected to preserve externally observable behavior even when the internal structure changes.

This notion is reflected in the kinds of behavioral deviations present in our dataset. In some cases, the difference is directly observable when executing the \texttt{main} method, for instance through changes in printed output. In many other cases, the deviation becomes evident only when invoking specific methods and checking their return values or the values they print. A smaller subset of bugs manifests through more subtle externally visible changes, such as altered method-call resolution, modified field access, changes in the accessible API, or the introduction of new exceptions. Our notion of behavior preservation is designed to capture not only output mismatches, but also failures that affect which method is invoked, whether a call throws an exception, or whether the visible program state observed through public operations remains the same.

The motivating examples in Section~\ref{sec:example} illustrate how this notion of equivalence is applied in practice. In the \textit{Push Down Method} example from Figure~\ref{fig:exampleOWC}, both versions compile and can be executed through the same public entry point, namely \texttt{C.main}. However, the observable output changes from \texttt{10} in the original program to \texttt{20} in the refactored program. Under our notion, this is a behavioral change because a client executing the same public operation observes a different printed value. Similarly, in the \textit{Extract Method} example from Figure~\ref{fig:exampleOWC11}, the original program executes \texttt{A.main} and prints \texttt{ok}, whereas the refactored program evaluates \texttt{x[0]} before entering the extracted method and therefore throws an \texttt{ArrayIndexOutOfBoundsException}. This also violates behavior preservation: the same client-level execution changes from normal termination with an observable output to abnormal termination with an exception. 

To make this notion operational, for each of the \bugsBC{} behavioral-change bugs in the dataset, we provide one JUnit test class containing a single test case that exposes the behavioral difference between the original and resulting programs. Importantly, the exact same test is compiled and executed unchanged against both program versions. A bug is considered a behavioral-change instance only when this shared test succeeds in revealing different outcomes across the two versions. These tests do not rely on Java reflection; instead, they interact with the programs only through ordinary Java constructs available to client code. This choice keeps the notion of observable behavior aligned with client-level usage scenarios.

The tests also provide a concrete view of the kinds of behavioral properties considered in our study. Many tests assert return values, such as checking that a method returns \texttt{1}, \texttt{0}, \texttt{10}, or another expected value. Others capture and compare standard output, which is necessary for programs whose externally visible behavior is expressed through printing. Several tests assert that no exception should be thrown, thereby capturing cases where a faulty refactoring introduces failures such as \texttt{ClassCastException}, or other unintended runtime exceptions. In a few cases, the tests check externally visible state after invoking a public method, such as verifying the value of a field that should have been updated according to the original behavior. 
Finally, the tests range from 8 to 23 LOC, with an average size of 15 LOC. This compact size suggests that, although the behavioral changes caused by faulty refactorings may be subtle, they can still be exposed by small and direct client-level tests in our dataset.
}

\subsection{Metrics}

To assess the \revision{accuracy and consistency} of foundation models in detecting compilation errors and behavioral changes, we adopt the accuracy and stability metrics presented in Table~\ref{tab:metrics}. These metrics~\cite{pass-k,atil2025nondeterminismdeterministicllmsettings} allow us to capture not only whether a model produces correct results, but also whether it does so consistently across multiple attempts, an \revision{important} property for tools \revision{intended to support} developers.

\begin{table}[t]
\centering
\caption{Accuracy and stability metrics used in our study.}
\label{tab:metrics}
\small
\begin{tabularx}{\linewidth}{p{2cm} p{2.8cm} X}
\hline
 & \textbf{Metric} & \textbf{Definition} \\
\hline
\multirow{2}{*}{Accuracy} 
& Mean Accuracy & Computes the rate of correct answers across \textit{k} responses for a question. The overall mean accuracy is the average over all questions. \\
& \textit{pass@\(k\)} & Binary metric that returns 1 if there is at least one correct answer across \textit{k} responses for a question, and 0 otherwise. The overall \textit{pass@\(k\)} is the average over all questions. Higher values indicate that the model is more likely to produce a correct response eventually. \\
\hline
\multirow{3}{*}{Stability} 
& Accuracy Spread & Computes the difference between the maximum and minimum mean accuracy across \textit{k} attempts over all questions. Lower values indicate that the model is more stable in accuracy across attempts. \\
& \textit{tar@\(k\)} & Binary metric that returns 1 if all \textit{k} answers for a question are identical, regardless of correctness, and 0 otherwise. The overall \textit{tar@\(k\)} is the average over all questions. Higher values indicate that the model produces more self-consistent responses across attempts. \\
& \textit{cons@\(k\)} & Binary metric that returns 1 if the unique most frequent answer, or consensus, across \textit{k} responses for a question is correct, and 0 otherwise. The overall \textit{cons@\(k\)} is the average over all questions. Higher values indicate that the model is more likely to produce a correct response consistently. \\
\hline
\end{tabularx}
\end{table}

Accuracy is measured through mean accuracy, which reflects the overall proportion of correct answers across responses, and \textit{pass@\(k\)}, which evaluates whether at least one correct answer emerges within \textit{k} attempts. While mean accuracy indicates the model’s average effectiveness, \textit{pass@\(k\)} highlights its potential to eventually provide a correct response under repeated queries, a scenario \revision{that may be useful} in automated pipelines. \revision{Because our benchmark contains only positive instances, there are no true negatives; therefore, in our setting, the reported accuracy numerically coincides with recall.}

Stability metrics complement these measures by examining how dependable model outputs are across different executions. Accuracy spread quantifies the gap between the best and worst results over multiple attempts, thereby reflecting the predictability of performance. \textit{tar@\(k\)} (Total Agreement Rate) measures the extent to which all responses for a question are identical, regardless of correctness, providing insights into the model's self-consistency. Finally, \textit{cons@\(k\)} evaluates whether the most frequent response across attempts aligns with the correct answer, thus combining correctness with consistency.

Together, these metrics provide a \revision{multi-dimensional} evaluation framework. Accuracy indicates whether models can identify the target transformations, while stability measures whether their predictions are consistent across runs. Considering both dimensions is important, since even accurate models may be difficult to use in developer-facing workflows if their behavior is inconsistent.

%Moreover, two authors of this article manually analyzed the responses of all models. In some cases, the models correctly identified that a transformation was invalid but provided an inaccurate explanation. Since a valid explanation is essential for effectively diagnosing refactoring bugs, we classified such responses as incorrect. In a few cases, disagreements arose during the classification process, which were subsequently discussed and resolved with the second author.

\subsection{Models}

The models evaluated include \gptoss{} and \gptfive{}, both developed by OpenAI. 
\revision{
\gptoss{} was chosen as a representative open-weight model that can be executed locally, offering potentially improved reproducibility, deployment flexibility, and lower-cost experimentation. In contrast, \gptfive{} was selected as a strong proprietary-model baseline, ranked highly in LLM Arena~\cite{llm-arena}. This contrast allows us to investigate an important practical trade-off between accessibility and performance, namely how far locally deployable \revisionTwo{open-weight} models can go for refactoring-bug detection and how they compare with stronger closed models in terms of effectiveness and stability.
}

We accessed \gptfive{} (\textsc{gpt-5.4-2026-03-05}) via its API, \revision{using the default settings, including a reasoning effort of \texttt{none}}, while \gptoss{} was executed locally using the Ollama framework in Python on a Mac Mini M4 Pro with 64GB of RAM. For \gptoss{}, we specified the model name and base URL, and used the default configurations of the LangChain Ollama API~\cite{ollama-setup}, with the exception of the temperature parameter, which was set to \revision{0.5} (see Section~\ref{sec:discussion-temp}). For \gptfive{}, we used the Python API with default settings, requiring only the API key~\cite{openai-api}. 
\revision{
We ran \gptoss{} and \gptfive{} five times each to assess prediction stability under repeated executions, since foundation models may vary even for the same input. We limited the evaluation to five runs per model to keep computational costs manageable. All analyses were conducted in April 2026.
}

\subsection{Correctness Assessment}
\label{sec:correctness-assessment}

\revision{
We assess each model response by comparing it against the ground truth established in Section~\ref{sec:dataset}. Since all \bugs{} instances in our dataset correspond to faulty refactorings, each instance is expected to be classified either as a behavioral change or as a compilation error. Therefore, a response that claims behavior preservation, i.e., \texttt{YES}, is always counted as incorrect in this benchmark. To make this process systematic and reproducible, we implemented a dedicated analysis tool that automatically evaluates each model answer according to the expected bug type and the corresponding validation criteria.
}

\revision{
For compilation-error instances, a response is considered correct only when the model classifies the transformation as \texttt{NO - COMPILATION ERROR}. This classification is checked against the dataset evidence showing that the original program compiles successfully whereas the resulting program does not. For behavioral-change instances, a response is considered correct only when the model classifies the transformation as \texttt{NO - BEHAVIOR CHANGE} and the JUnit test provided in the model response constitutes valid executable evidence of the behavioral difference. Specifically, our tool compiles the model-provided test against both the original and resulting programs using \openjdk{} and records the corresponding compiler logs. It then executes the same test on both versions and records the execution results. The response is counted as correct only if the same model-provided test compiles successfully in both versions and exposes different outcomes across them, that is, it succeeds on one version and fails on the other. 
}

\subsection{Prompt}
\label{sec:methodology-prompts}

We adopt a zero-shot prompting strategy~\cite{prompts,prompt-techniques,zero-shot-prompt}, in which the model is asked to solve the task without task-specific fine-tuning or example-based demonstrations. In addition, we employed MetaPrompting~\cite{hou-metaprompting} to help refine the prompt design. More specifically, we iteratively asked \gptfive{} to improve an initial prompt designed to assess whether a refactoring introduces compilation errors or behavioral changes. Through this process, we obtained a revised prompt with a more structured output format and clearer instructions regarding compilation correctness and behavioral preservation. 

\revision{
The final prompt takes as input the original program before refactoring (\texttt{code1}) and the resulting program after refactoring (\texttt{code2}), and asks the model to check two conditions: (i) whether the resulting program compiles successfully, and (ii) whether the transformation preserves behavior according to the notion of behavior preservation defined in Section~\ref{sec:behavior-preservation}. To support automatic processing, the model must return valid JSON with three fields: \texttt{verdict}, \texttt{explanation}, and \texttt{junit\_test}. The \texttt{verdict} must be exactly one of \texttt{YES}, \texttt{NO - COMPILATION ERROR}, or \texttt{NO - BEHAVIOR CHANGE}; the \texttt{explanation} must provide a brief evidence-based justification; and \texttt{junit\_test} must be \texttt{null} unless the verdict is \texttt{NO - BEHAVIOR CHANGE}. In that case, the model must generate one deterministic JUnit test, as a complete Java source file, that compiles and runs unchanged against both versions and exposes the behavioral difference. This JSON-based design makes the output easier to process automatically and reduces ambiguities associated with free-form answers. Next, we present the prompt used in our experiments.
}

\begin{tcolorbox}[
  breakable,
  colback=gray!5,
  colframe=black!60,
  boxrule=0.5pt,
  arc=2pt,
  left=6pt,right=6pt,top=6pt,bottom=6pt
]
\footnotesize
\noindent Consider the following initial program: \\

\noindent \texttt{<CODE1} \\
\noindent \texttt{\{code1\}} \\
\noindent \texttt{CODE1>} \\

\noindent After applying a refactoring, it yields the following resulting program: \\

\noindent \texttt{<CODE2} \\
\noindent \texttt{\{code2\}} \\
\noindent \texttt{CODE2>} \\

\noindent Check the following: \\

\noindent 1. Compilation correctness: \\
\noindent The resulting program compiles successfully, with no syntax, type, name-resolution, import, or linkage errors. \\

\noindent 2. Behavioral equivalence: \\
\noindent Ignore methods not present in both programs; compare only methods with the same signature that are public in both versions. For those common public methods, the initial and resulting programs preserve the same observable behavior, including return values, printed output, thrown exceptions, and externally visible state changes.
\\

\noindent Return ONLY valid JSON. Do not use markdown. Do not add any text before or after the JSON. \\

\noindent Use exactly this schema: \\
\noindent \texttt{\{} \\
\noindent \texttt{\ \ "verdict": "YES | NO - COMPILATION ERROR | NO - BEHAVIOR CHANGE",} \\
\noindent \texttt{\ \ "explanation": "1-4 sentences",} \\
\noindent \texttt{\ \ "junit\_test": null} \\
\noindent \texttt{\}} \\

\noindent Rules: \\
\noindent - \texttt{"verdict"} must be exactly one of: \\
\noindent \texttt{"YES"} \\
\noindent \texttt{"NO - COMPILATION ERROR"} \\
\noindent \texttt{"NO - BEHAVIOR CHANGE"} \\
\noindent - \texttt{"explanation"} must briefly justify the answer with concrete evidence. \\
\noindent - \texttt{"junit\_test"} must be \texttt{null} unless the verdict is \texttt{"NO - BEHAVIOR CHANGE"}. \\
\noindent - If \texttt{"junit\_test"} is not \texttt{null}, it must contain exactly one deterministic JUnit test as a complete Java source file. \\
\noindent - The content of \texttt{"junit\_test"} must be directly copyable into a Java IDE such as VS Code. \\
\noindent - The exact same generated test code must compile and run against both \texttt{code1} and \texttt{code2} without any modification. \\
\noindent - The test must expose a behavioral difference by producing one outcome on \texttt{code1} and a different outcome on \texttt{code2}. \\
\noindent - The test must include all necessary imports, exactly one public test class, and exactly one test method. \\
\noindent - The test must be minimal, concrete, deterministic, and focused on the changed public behavior. \\
\noindent - The test must not rely on randomness, time, concurrency, network, filesystem, or environment-specific behavior. \\
\noindent - Do not invent compilation errors. \\
\noindent - Do not claim behavior change unless you can point to a specific differing return value, output, exception, or externally visible state. \\
\noindent - Do not include markdown fences such as \texttt{java} inside \texttt{"junit\_test"}. \\
\noindent - Write \texttt{"junit\_test"} as plain Java code text. \\

\noindent Return JSON only.
\end{tcolorbox}

\revision{
In this prompt, \texttt{code1} and \texttt{code2} denote the program versions before and after the refactoring, respectively. For example, these placeholders correspond to the programs shown in Figures~\ref{fig:exampleOWC}(a) and~\ref{fig:exampleOWC}(b), respectively.
}

\subsection{Metamorphic Testing}
\label{sec:mt}

A key threat to validity when evaluating foundation models is data contamination~\cite{sallou2024breaking}, that is, the possibility that a model has been exposed during training to identical or highly similar examples, which may artificially inflate performance. \revisionTwo{To assess robustness under controlled behavior-preserving perturbations, rather than to rule out memorization or contamination, we applied metamorphic testing (MT)~\cite{metamorphic-testing,metamorphic-testing-2} to all \bugs{} transformations in our benchmark.} The main idea is to generate behaviorally equivalent variants of the original programs while altering their lexical and structural form. This allows us to assess whether the models produce similar predictions when the same underlying refactoring bug is presented through a syntactically different original program.

\revision{
For each bug instance, we applied a metamorphic transformation to the original program before evaluating the corresponding refactoring scenario. The resulting program was kept unchanged. This setup tests whether the model can still identify the same refactoring bug when the original program is presented through a behavior-preserving variant with a different lexical or structural form.
The metamorphic transformation operators were intended to preserve behavior under the equivalence notion adopted in this study, as defined in Section~\ref{sec:behavior-preservation}. The operators used in this analysis do not change existing control flow, method-call resolution, return expressions, printed output, exception behavior, or public operations exercised by the tests. Instead, they introduce behaviorally inert modifications, such as unused declarations, comments, imports, and auxiliary classes. Therefore, these transformations are expected not to change whether an instance is classified as a behavioral change or as a compilation error.

We implemented six metamorphic transformation operators on top of \textsc{Spoon}~\cite{spoon}: \textit{AddFieldOperator} (AF), \textit{CommentsOperator} (CO), \textit{InnerClassOperator} (IC), \textit{JavaImportOperator} (JI), \textit{LocalVariableDeclarationOperator} (LVD), and \textit{TopLevelClassOperator} (TLC). These operators introduce intended behavior-preserving changes such as adding unused fields, injecting comments, creating auxiliary inner or top-level classes, inserting unused imports, and declaring dead local variables. Table~\ref{tab:mt-examples} shows representative examples of these transformations applied to the original program.
}

\begin{table*}[t]
\centering
\caption{Examples of metamorphic transformation operators (\textit{AddFieldOperator} (AF), \textit{CommentsOperator} (CO), \textit{InnerClassOperator} (IC), \textit{JavaImportOperator} (JI), \textit{LocalVariableDeclarationOperator} (LVD), and \textit{TopLevelClassOperator} (TLC)) applied to the original program to generate behavior-preserving variants.}
\label{tab:mt-examples}
\small
\begin{tabular}{p{0.10\textwidth}p{0.40\textwidth}p{0.42\textwidth}}
\hline
\textbf{Operator} & \textbf{Before} & \textbf{After} \\
\hline

AF &
\begin{lstlisting}[style=refjava]
class A {
  int m() { return 1; }
}
\end{lstlisting}
&
\begin{lstlisting}[style=refjava]
class A {
  int field = 42;
  int m() { return 1; }
}
\end{lstlisting}
\\
\hline

CO &
\begin{lstlisting}[style=refjava]
class A {
  int m() { return 1; }
}
\end{lstlisting}
&
\begin{lstlisting}[style=refjava]
class A {
  // comment
  int m() { return 1; }
}
\end{lstlisting}
\\
\hline

JI &
\begin{lstlisting}[style=refjava]
class A {
  int m() { return 1; }
}
\end{lstlisting}
&
\begin{lstlisting}[style=refjava]
import java.util.List;

class A {
  int m() { return 1; }
}
\end{lstlisting}
\\
\hline

LVD &
\begin{lstlisting}[style=refjava]
class A {
  int m() { return 1; }
}
\end{lstlisting}
&
\begin{lstlisting}[style=refjava]
class A {
  int m() {
    int local = 7;
    return 1;
  }
}
\end{lstlisting}
\\
\hline

IC &
\begin{lstlisting}[style=refjava]
class A {
  int m() { return 1; }
}
\end{lstlisting}
&
\begin{lstlisting}[style=refjava]
class A {
  class Aux {
    int value = 0;
  }
  int m() { return 1; }
}
\end{lstlisting}
\\
\hline

TLC &
\begin{lstlisting}[style=refjava]
class A {
  int m() { return 1; }
}
\end{lstlisting}
&
\begin{lstlisting}[style=refjava]
class A {
  int m() { return 1; }
}

class Aux {
  int value = 0;
}
\end{lstlisting}
\\
\hline

\end{tabular}
\end{table*}

\revision{
Although Table~\ref{tab:mt-examples} shows simplified examples, the actual transformations are randomized. Both the choice of operator and the internal choices within each operator vary across instances. Depending on the operator, this includes random variation in identifiers, literals, comments, imports, declared types, field names, class names, and generated class structures. For example, AF randomly chooses the type and initial value of the injected field; CO randomly selects among several comments; JI randomly inserts one among several imports; and LVD randomly varies both the declared type and assigned literal. Similarly, IC and TLC generate auxiliary classes with randomized names, field names, field types, and default values. As a result, even when two instances are transformed by the same operator, the transformed original programs may differ at both lexical and structural levels.
To further strengthen the validity of the metamorphic analysis, we checked the generated variants through compilation and test execution. For the models evaluated in this analysis, we stored the compiler logs for the transformed original programs and the corresponding resulting programs, the compiler logs for the JUnit tests provided in the model responses, and the JUnit execution results obtained by running those tests on both program versions. 

Table~\ref{tab:mt-transformations} summarizes the number of transformed instances per operator. The distribution is reasonably balanced, with all six operators being applied dozens of times. 
An additional minor contribution of this study is the implementation infrastructure itself. We developed the metamorphic transformation operators as a reusable Java-based framework, including implementations for AF, CO, IC, JI, LVD, and TLC. This framework can be used or extended by future studies interested in generating intended behavior-preserving program variants for robustness assessment in refactoring analysis and foundation-model evaluation.
}

\begin{table}[t]
\centering
\caption{Metamorphic operators applied to the \bugs{} instances.}
\label{tab:mt-transformations}
\small
\begin{tabular}{llr}
\hline
\textbf{Operator} & \textbf{Abbrev.} & \textbf{Instances} \\
\hline
\textit{AddFieldOperator}                  & AF  & 32 \\
\textit{CommentsOperator}                  & CO  & 40 \\
\textit{InnerClassOperator}                & IC  & 37 \\
\textit{JavaImportOperator}                & JI  & 37 \\
\textit{LocalVariableDeclarationOperator}  & LVD & 45 \\
\textit{TopLevelClassOperator}             & TLC & 35 \\
\hline
\textbf{Total} & & \bugs{} \\
\hline
\end{tabular}
\end{table}

\revisionTwo{
Therefore, the metamorphic analysis should be interpreted as a robustness evaluation under the specific behavior-preserving perturbations implemented in this study. Because the applied operators introduce mostly inert modifications, the transformed programs may remain recognizable to a model that has previously seen the original code or closely related variants, since semantic similarity persists at the token and structural levels. Consequently, the results should not be interpreted as strong evidence against memorization, data contamination, or higher-order clone recognition.
}

\section{Results}
\label{sec:results}

Next we answer our research questions.

\subsection{RQ$_{1}$. To what extent can an \revisionTwo{open-weight} foundation model (\gptoss{}) detect refactoring bugs, including compilation errors and behavioral changes?}
\label{sec:results-rq1-attempts}

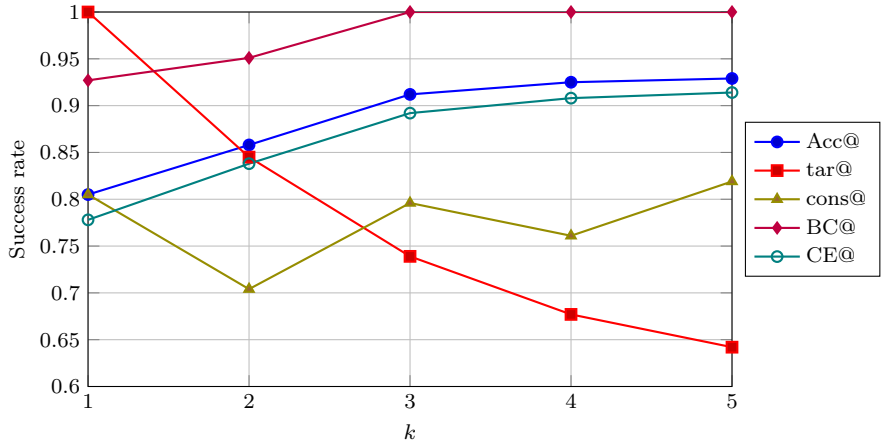
\begin{figure}[htbp]
    \centering
    \begin{tikzpicture}
        \begin{axis}[
            width=0.85\textwidth,
            height=0.55\textwidth,
            xlabel={$k$},
            ylabel={Success rate},
            xmin=1, xmax=5,
            ymin=0.60, ymax=1.00,
            xtick={1,2,3,4,5},
            ytick={0.60,0.65,0.70,0.75,0.80,0.85,0.90,0.95,1.00},
            grid=major,
            legend style={at={(1.02,0.5)}, anchor=west},
            legend cell align={left}
        ]

        % Acc@
        \addplot+[thick, mark=*, color=blue] coordinates {
            (1,0.805) (2,0.858) (3,0.912) (4,0.925) (5,0.929)
        };
        \addlegendentry{Acc@}

        % tar@
        \addplot+[thick, mark=square*, color=red] coordinates {
            (1,1.000) (2,0.845) (3,0.739) (4,0.677) (5,0.642)
        };
        \addlegendentry{tar@}

        % cons@
        \addplot+[thick, mark=triangle*, color=olive] coordinates {
            (1,0.805) (2,0.704) (3,0.796) (4,0.761) (5,0.819)
        };
        \addlegendentry{cons@}

        % BC@
        \addplot+[thick, mark=diamond*, color=purple] coordinates {
            (1,0.927) (2,0.951) (3,1.000) (4,1.000) (5,1.000)
        };
        \addlegendentry{BC@}

        % CE@
        \addplot+[thick, mark=o, color=teal] coordinates {
            (1,0.778) (2,0.838) (3,0.892) (4,0.908) (5,0.914)
        };
        \addlegendentry{CE@}

        \end{axis}
    \end{tikzpicture}
    \caption{\gptoss{} results across five repeated attempts. Acc@ = cumulative overall success rate considering whether an instance is solved in at least one of the first $k$ attempts, tar@ = agreement in binary correctness across the first $k$ attempts, cons@ = correctness of the strict-majority outcome across the first $k$ attempts, BC@ = cumulative success rate on behavioral-change cases, and CE@ = cumulative success rate on compilation-error cases. \revision{CE = Compilation Error, and BC = Behavioral Change.}}
    \label{fig:results-gpt-oss-attempts}
\end{figure}

\revision{
Across individual runs, \gptoss{} exhibits relatively stable behavior in this setting, although non-negligible variation remains even under a fixed configuration. Overall accuracy ranges from $0.757$ to $0.805$. The mean overall accuracy across the five runs is $0.784$.
Figure~\ref{fig:results-gpt-oss-attempts} summarizes five complementary metrics over the first $k$ attempts: cumulative overall success rate (Acc@), agreement in binary correctness across attempts (tar@), strict-majority correctness (cons@), cumulative success on behavioral-change cases (BC@), and cumulative success on compilation-error cases (CE@).
The model is consistently stronger on behavioral-change instances than on compilation-error instances in our experiments. BC accuracy ranges from $0.854$ to $0.927$, whereas CE accuracy ranges from $0.724$ to $0.778$. \gptoss{} performs better on BC than on CE in every attempt. The strongest overall result (Attempt~1) combines the highest BC accuracy ($0.927$) with one of the two highest CE accuracies ($0.778$).

Considering cumulative success across attempts reveals that repeated sampling increases cumulative coverage. With only the first attempt, \gptoss{} correctly handles 182 out of \bugs{} instances. Acc@ increases to $0.858$ after two attempts, $0.912$ after three attempts, $0.925$ after four attempts, and $0.929$ after all five attempts. 
The cumulative curves also highlight a clear contrast between BC and CE. For BC, cumulative success increases from $0.927$ with one attempt to $0.951$ with two attempts and reaches $1.000$ with three attempts. For CE, cumulative success increases more gradually, from $0.778$ with one attempt to $0.838$, $0.892$, $0.908$, and $0.914$ after two, three, four, and five attempts, respectively. This indicates that repeated sampling is effective in this setting for recovering BC cases that fail in individual runs, but it does not eliminate the hardest CE failures.

The agreement and majority-based metrics provide a complementary view of stability. The tar@ curve decreases from $1.000$ for a single attempt to $0.845$, $0.739$, $0.677$, and $0.642$ when considering the first two, three, four, and five attempts, respectively. This decline shows that a notable fraction of instances changes correctness status across runs, even though many of these fluctuations are beneficial for cumulative coverage. The cons@ curve is not monotonic because it depends on the strict-majority outcome among the first $k$ attempts. It starts at $0.805$, decreases to $0.704$ for two attempts due to unresolved ties, and then reaches $0.796$, $0.761$, and $0.819$ for three, four, and five attempts.

The BC subset is largely attempt-sensitive. No BC instance is misclassified in all five runs. However, a few cases recur across attempts. IDs~72 (\textit{Change Method Signature}) and~281 (\textit{Rename Field}) are the most difficult BC cases, each failing in three out of five attempts. ID~72 fails either because the generated test does not compile or because it does not expose the behavioral change, suggesting instability in test generation. ID~281 repeatedly fails because the model states that behavior was preserved, indicating a recurring incorrect semantic verdict. BC errors are not concentrated in a single structural category, but instead arise from transformations that require sensitivity to subtle changes in data flow, conditions, name binding, or interface contracts.

The CE subset, in contrast, contains a more persistent core of difficult instances. Sixteen CE IDs are misclassified in all five attempts.
% : 15, 94, 106, 111, 122, 123, 131, 154, 192, 203, 225, 326, 345, 349, 397, and~398
These cases represent recurring failure modes rather than occasional fluctuations. In these persistent CE failures, the model repeatedly assigns a semantic verdict, stating either that behavior changed or that behavior was preserved, when the correct label is compilation error.
The persistent CE failures are concentrated in transformations that alter declarations, member placement, signatures, or class hierarchies. The CE cases misclassified in all five attempts include \textit{Push Down Method} (IDs~15 and~345), \textit{Move Method} (IDs~94 and~131), \textit{Inline Variable} (IDs~106 and~326), \textit{Rename Method} (IDs~111 and~397), \textit{Introduce Constant} (IDs~122 and~123), \textit{Pull Up Method} (IDs~154, 192, and~398), \textit{Move Class} (ID~203), \textit{Pull Up Field} (ID~225), and \textit{Inline Method} (ID~349).
This indicates that the observed difficulty appears to be driven less by raw frequency and more by the semantic and structural challenges posed by transformations that affect name resolution, inheritance, member visibility, and compilation validity.
}

\begin{tcolorbox}[
    colback=gray!8,
    colframe=black!70,
    title={RQ$_{1}$ Answer},
    fonttitle=\bfseries,
    sharp corners,
    boxrule=0.6pt
]
\noindent
\gptoss{} shows moderate stability in this setting, with stronger and more recoverable performance on behavioral-change cases than on compilation-error cases. Repeated sampling improves cumulative coverage to 92.9\%, indicating that multiple attempts can recover some missed bugs. However, strict-majority aggregation is less effective because several instances alternate between correct and incorrect outcomes. Overall, the main limitation of \gptoss{} lies in persistent CE cases involving name resolution, inheritance, member visibility, and compilation validity.
\end{tcolorbox}

\subsection{RQ$_{2}$. To what extent can a proprietary foundation model (\gptfive{}) detect refactoring bugs, including compilation errors and behavioral changes?}
\label{sec:results-rq2}

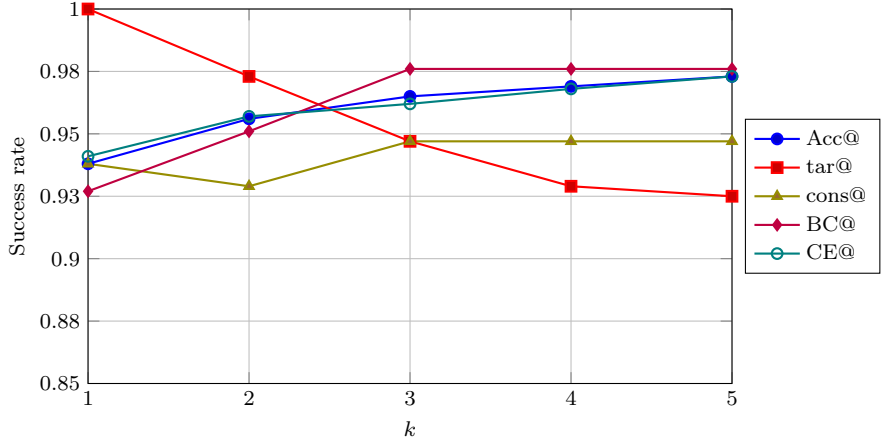
\begin{figure}[htbp]
    \centering
    \begin{tikzpicture}
        \begin{axis}[
            width=0.85\textwidth,
            height=0.55\textwidth,
            xlabel={$k$},
            ylabel={Success rate},
            xmin=1, xmax=5,
            ymin=0.85, ymax=1.00,
            xtick={1,2,3,4,5},
            ytick={0.85,0.875,0.90,0.925,0.95,0.975,1.00},
            grid=major,
            legend style={at={(1.02,0.5)}, anchor=west},
            legend cell align={left}
        ]

        % Acc@
        \addplot+[thick, mark=*, color=blue] coordinates {
            (1,0.938) (2,0.956) (3,0.965) (4,0.969) (5,0.973)
        };
        \addlegendentry{Acc@}

        % tar@
        \addplot+[thick, mark=square*, color=red] coordinates {
            (1,1.000) (2,0.973) (3,0.947) (4,0.929) (5,0.925)
        };
        \addlegendentry{tar@}

        % cons@
        \addplot+[thick, mark=triangle*, color=olive] coordinates {
            (1,0.938) (2,0.929) (3,0.947) (4,0.947) (5,0.947)
        };
        \addlegendentry{cons@}

        % BC@
        \addplot+[thick, mark=diamond*, color=purple] coordinates {
            (1,0.927) (2,0.951) (3,0.976) (4,0.976) (5,0.976)
        };
        \addlegendentry{BC@}

        % CE@
        \addplot+[thick, mark=o, color=teal] coordinates {
            (1,0.941) (2,0.957) (3,0.962) (4,0.968) (5,0.973)
        };
        \addlegendentry{CE@}

        \end{axis}
    \end{tikzpicture}
    \caption{\gptfive{} results across five repeated attempts. Acc@ = cumulative overall success rate considering whether an instance is solved in at least one of the first $k$ attempts, tar@ = agreement in correctness and error status across the first $k$ attempts, cons@ = correctness of the strict-majority outcome across the first $k$ attempts, BC@ = cumulative success rate on behavioral-change cases, and CE@ = cumulative success rate on compilation-error cases. \revision{CE = Compilation Error, and BC = Behavioral Change.}}
    \label{fig:gpt-results}
\end{figure}

\revision{

\gptfive{} achieves high single-attempt performance in this benchmark and remains stable across repeated executions. In individual runs, overall accuracy ranges from $0.938$ to $0.947$, with a spread of only $0.009$. This indicates that, in our setting, \gptfive{} already provides high performance without requiring multiple attempts.
Figure~\ref{fig:gpt-results} reports five complementary metrics for $k$: cumulative overall success rate (Acc@), agreement in correctness and error status across attempts (tar@), strict-majority correctness (cons@), cumulative success on behavioral-change cases (BC@), and cumulative success on compilation-error cases (CE@).
The cumulative results show that additional attempts provide modest but consistent gains. Acc@ increases from $0.938$ at $k{=}1$ to $0.956$ at $k{=}2$, $0.965$ at $k{=}3$, $0.969$ at $k{=}4$, and $0.973$ at $k{=}5$. Across all five runs, \gptfive{} correctly handles 220 out of \bugs{} instances at least once. These results show that repeated sampling still improves coverage, but most of the model's effectiveness in this benchmark is already achieved in the first attempt.
By type, \gptfive{} performs slightly better on compilation-error instances than on behavioral-change instances. Across individual runs, BC accuracy ranges from $0.878$ to $0.927$, whereas CE accuracy ranges from $0.941$ to $0.957$. The cumulative curves follow the same pattern. BC@ increases from $0.927$ at $k{=}1$ to $0.951$ at $k{=}2$, reaches $0.976$ at $k{=}3$, and remains at $0.976$ through $k{=}5$. CE@ increases from $0.941$ at $k{=}1$ to $0.957$, $0.962$, $0.968$, and $0.973$ for $k{=}2$ through $k{=}5$. 

The agreement and majority-based metrics further support the model's stability in this setting. The tar@ curve decreases from $1.000$ for a single attempt to $0.973$, $0.947$, $0.929$, and $0.925$ when considering the first two, three, four, and five attempts, respectively. This decline is expected because exact agreement becomes harder as more attempts are considered, but the values remain high. The cons@ curve also remains high: it starts at $0.938$, decreases slightly to $0.929$ for two attempts due to unresolved ties, and reaches $0.947$ for $k{=}3$, $k{=}4$, and $k{=}5$. This indicates that the majority answer is usually correct even when individual attempts differ.

For BC, there are 20 total errors across the five attempts. The most common failure mode is an incorrect semantic verdict that behavior is preserved, accounting for 10 cases. In 4 cases, the model incorrectly indicates a compilation error. Test-generation failures account for the remaining cases: 5 errors occur because the generated test does not compile, and 1 occurs because the generated test does not expose the behavioral change. 
For CE, there are 46 total errors across the five attempts. The dominant failure mode is assigning a semantic verdict to an uncompilable transformation: in 30 cases, the model indicates a behavior change when the correct label is compilation error, while in 16 cases it states that behavior is preserved. This suggests that the hardest CE failures are not caused primarily by output instability, but by treating syntactically invalid transformations as if they were semantically analyzable.

In the BC subset, only ID~112 (\textit{Rename Method}) is misclassified in all five attempts, and it repeatedly fails because the model states that behavior was preserved. ID~281 (\textit{Rename Field}) fails in four attempts, alternating between indicating a compilation error and stating that behavior was preserved. ID~361 (\textit{Rename Field}) fails in three attempts because the generated test does not compile. 
In the CE subset, five IDs are misclassified in all five attempts: 104 (\textit{Inline Variable}), 108 (\textit{Inline Variable}), 215 (\textit{Move Method}), 282 (\textit{Inline Variable}), and~360 (\textit{Push Down Field}). Additional CE IDs recur in four out of five attempts. The recurring CE errors again reflect semantic misclassification: the model either indicates a behavior change or states that behavior was preserved when the correct answer is compilation error.
Figure~\ref{fig:heatmap-refactoring-types} summarizes this analysis by comparing \gptfive{} and \gptoss{} across the 10 most frequent refactoring types in the dataset. 
}

\begin{figure*}[t]
\centering
\begin{tikzpicture}[font=\small]

% ---------- parameters ----------
\def\cellw{1.45}
\def\cellh{0.62}
\def\leftx{0}
\def\rightx{4.6}
\def\topy{0}

% ---------- helper macro ----------
% #1 x, #2 y, #3 value in [0,1], #4 label
\newcommand{\heatcell}[4]{%
    \pgfmathsetmacro{\shade}{int(100*#3)}
    \pgfmathparse{#3 >= 0.75 ? 1 : 0}
    \ifnum\pgfmathresult=1
        \def\textcol{white}
    \else
        \def\textcol{black}
    \fi
    \fill[myblue!\shade!white] (#1,#2) rectangle ++(\cellw,-\cellh);
    \draw[white, line width=0.8pt] (#1,#2) rectangle ++(\cellw,-\cellh);
    \node[text=\textcol] at ($(#1,#2)+(.5*\cellw,-.5*\cellh)$) {#4};
}

% empty cell with dash
\newcommand{\emptycell}[3]{%
    \fill[gray!12] (#1,#2) rectangle ++(\cellw,-\cellh);
    \draw[white, line width=0.8pt] (#1,#2) rectangle ++(\cellw,-\cellh);
    \node[text=black!70] at ($(#1,#2)+(.5*\cellw,-.5*\cellh)$) {#3};
}

% ---------- panel titles ----------
\node[font=\bfseries] at (\leftx+1.45,-0.35) {\gptoss{}};
\node[font=\bfseries] at (\rightx+1.45,-0.35) {\gptfive{}};

% ---------- column labels ----------
\node[font=\bfseries] at (\leftx+0.5*\cellw,-0.8) {CE};
\node[font=\bfseries] at (\leftx+1.5*\cellw,-0.8) {BC};
\node[font=\bfseries] at (\rightx+0.5*\cellw,-0.8) {CE};
\node[font=\bfseries] at (\rightx+1.5*\cellw,-0.8) {BC};

% ---------- y labels ----------
\node[anchor=east] at (\leftx-0.1,-1.30) {\textit{Move Method} (n=32)};
\node[anchor=east] at (\leftx-0.1,-1.92) {\textit{Inline Method} (n=20)};
\node[anchor=east] at (\leftx-0.1,-2.54) {\textit{Pull Up Method} (n=18)};
\node[anchor=east] at (\leftx-0.1,-3.16) {\textit{Extr. Local Variable} (n=15)};
\node[anchor=east] at (\leftx-0.1,-3.78) {\textit{Rename Method} (n=14)};
\node[anchor=east] at (\leftx-0.1,-4.40) {\textit{Change Meth. Sig.} (n=13)};
\node[anchor=east] at (\leftx-0.1,-5.02) {\textit{Extract Method} (n=12)};
\node[anchor=east] at (\leftx-0.1,-5.64) {\textit{Move Class} (n=11)};
\node[anchor=east] at (\leftx-0.1,-6.26) {\textit{Inline Variable} (n=10)};
\node[anchor=east] at (\leftx-0.1,-6.88) {\textit{Push Down Method} (n=6)};
\node[anchor=east] at (\leftx-0.1,-7.50) {\textit{Others} (n=75)};

% ---------- left heatmap: GPT-OSS ----------
\heatcell{\leftx}{-1.00}{0.79}{0.79}
\heatcell{\leftx+\cellw}{-1.00}{0.67}{0.67}

\heatcell{\leftx}{-1.62}{0.73}{0.73}
\heatcell{\leftx+\cellw}{-1.62}{1.00}{1.00}

\heatcell{\leftx}{-2.24}{0.69}{0.69}
\heatcell{\leftx+\cellw}{-2.24}{1.00}{1.00}

\heatcell{\leftx}{-2.86}{1.00}{1.00}
\heatcell{\leftx+\cellw}{-2.86}{1.00}{1.00}

\heatcell{\leftx}{-3.48}{0.50}{0.50}
\heatcell{\leftx+\cellw}{-3.48}{1.00}{1.00}

\heatcell{\leftx}{-4.10}{0.91}{0.91}
\heatcell{\leftx+\cellw}{-4.10}{0.50}{0.50}

\heatcell{\leftx}{-4.72}{0.91}{0.91}
\heatcell{\leftx+\cellw}{-4.72}{1.00}{1.00}

\heatcell{\leftx}{-5.34}{0.82}{0.82}
\emptycell{\leftx+\cellw}{-5.34}{--}

\heatcell{\leftx}{-5.96}{0.80}{0.80}
\emptycell{\leftx+\cellw}{-5.96}{--}

\heatcell{\leftx}{-6.58}{0.00}{0.00}
\heatcell{\leftx+\cellw}{-6.58}{1.00}{1.00}

\heatcell{\leftx}{-7.20}{0.80}{0.80}
\heatcell{\leftx+\cellw}{-7.20}{0.93}{0.93}

% ---------- right heatmap: GPT-5 ----------
\heatcell{\rightx}{-1.00}{0.93}{0.93}
\heatcell{\rightx+\cellw}{-1.00}{1.00}{1.00}

\heatcell{\rightx}{-1.62}{1.00}{1.00}
\heatcell{\rightx+\cellw}{-1.62}{1.00}{1.00}

\heatcell{\rightx}{-2.24}{0.88}{0.88}
\heatcell{\rightx+\cellw}{-2.24}{1.00}{1.00}

\heatcell{\rightx}{-2.86}{1.00}{1.00}
\heatcell{\rightx+\cellw}{-2.86}{1.00}{1.00}

\heatcell{\rightx}{-3.48}{0.90}{0.90}
\heatcell{\rightx+\cellw}{-3.48}{0.75}{0.75}

\heatcell{\rightx}{-4.10}{1.00}{1.00}
\heatcell{\rightx+\cellw}{-4.10}{1.00}{1.00}

\heatcell{\rightx}{-4.72}{1.00}{1.00}
\heatcell{\rightx+\cellw}{-4.72}{1.00}{1.00}

\heatcell{\rightx}{-5.34}{1.00}{1.00}
\emptycell{\rightx+\cellw}{-5.34}{--}

\heatcell{\rightx}{-5.96}{0.70}{0.70}
\emptycell{\rightx+\cellw}{-5.96}{--}

\heatcell{\rightx}{-6.58}{1.00}{1.00}
\heatcell{\rightx+\cellw}{-6.58}{1.00}{1.00}

\heatcell{\rightx}{-7.20}{0.95}{0.95}
\heatcell{\rightx+\cellw}{-7.20}{0.86}{0.86}

% ---------- x-axis labels ----------
% \node[font=\bfseries] at (\leftx+1.45,-8.15) {Bug type};
% \node[font=\bfseries] at (\rightx+1.45,-8.15) {Bug type};

\end{tikzpicture}
\caption{Heatmap of model accuracy across the 10 most frequent refactoring types in the dataset, separated into compilation-error (CE) and behavioral-change (BC) instances. All remaining refactoring categories are grouped into \textit{Others}.}
\label{fig:heatmap-refactoring-types}
\end{figure*}

\begin{tcolorbox}[
    colback=gray!8,
    colframe=black!70,
    title={RQ$_{2}$ Answer},
    fonttitle=\bfseries,
    sharp corners,
    boxrule=0.6pt
]
\noindent
\gptfive{} is effective and stable at detecting reported refactoring bugs in this benchmark across repeated attempts. Repeated sampling improves cumulative coverage from 93.8\% to 97.3\%, but most gains are achieved in the first attempt, suggesting that repeated attempts provide limited additional benefit. The remaining failures are concentrated in a small number of persistent cases: BC failures mainly involve incorrect semantic verdicts or test-generation failures, whereas CE failures are dominated by transformations that affect bindings, member resolution, and structural relationships.
\end{tcolorbox}

\subsection{RQ$_{3}$. To what extent does the accuracy of these models remain stable under the semantics-preserving metamorphic transformations evaluated in this study?}
\label{sec:results-rq3}

\revision{
To assess robustness under potential training-data contamination concerns, we conducted a metamorphic testing analysis. We applied intended semantics-preserving source-code transformations to the input programs and re-evaluated both models on the transformed instances. In this analysis, the original setting corresponds to the first execution of each model, whereas the metamorphic-testing setting corresponds to the execution on the transformed programs. \revisionTwo{This analysis does not rule out memorization or contamination, because the transformed programs may remain semantically, structurally, and lexically similar to the original inputs. Instead, it evaluates whether model predictions remain stable under the specific behavior-preserving perturbations implemented in this study.}
}

Table~\ref{tab:mt-performance} summarizes the results before and after metamorphic testing. Overall, the metamorphic transformations had only limited impact on both models. For \gptoss{}, overall accuracy changes only slightly, from $0.805$ in the original setting to $0.801$ under metamorphic testing.  For \gptfive{}, the overall effect is similarly small: accuracy increases from $0.938$ to $0.942$. BC accuracy decreases from $0.927$ to $0.902$, whereas CE accuracy increases from $0.941$ to $0.951$. These results indicate that the predictions of both models are stable under the applied metamorphic transformations.

\begin{table}[t]
\centering
\caption{Performance before and after metamorphic testing. CE = Compilation Error; BC = Behavioral Change.}
\label{tab:mt-performance}
\small
\begin{tabular}{llccc}
\hline
\textbf{Model} & \textbf{Setting} & \textbf{Overall Acc.} & \textbf{BC Acc.} & \textbf{CE Acc.} \\
\hline
\gptoss{}  & Original             & 0.805 & 0.927 & 0.778 \\
\gptoss{}  & Metamorphic testing  & 0.801 & 0.951 & 0.768 \\
\hline
\gptfive{} & Original             & 0.938 & 0.927 & 0.941 \\
\gptfive{} & Metamorphic testing  & 0.942 & 0.902 & 0.951 \\
\hline
\end{tabular}
\end{table}

A closer inspection of the error IDs shows that metamorphic testing does not completely change which instances are difficult. Instead, it preserves a meaningful portion of the hard cases observed in the original execution, especially in the CE subset. For \gptoss{}, the BC overlap between the original and metamorphic settings is empty. In the CE subset, 22 IDs are misclassified in both settings. For \gptfive{}, the overlap is even more concentrated. In the BC subset, IDs~112 and~281 are misclassified in both the original and metamorphic settings. In the CE subset, 8 of the metamorphic CE errors were also present in the original run: IDs~104, 108, 215, 278, 282, 360, 398, and~401. This overlap suggests that metamorphic testing preserves many of the hardest compilation-error cases, especially for \gptfive{}. 

We also analyzed whether certain metamorphic operators (see Table~\ref{tab:mt-examples}) were associated with more errors than others. Table~\ref{tab:mt-operators} summarizes the distribution of metamorphic-testing errors by operator. For \gptoss{}, the MT errors are spread across all six operators, suggesting that no single transformation type fully explains the failures. In absolute terms, \textit{InnerClassOperator} accounts for the largest number of CE errors, followed by \textit{JavaImportOperator}, \textit{CommentsOperator}, and \textit{LocalVariableDeclarationOperator}. In the BC subset, \gptoss{} produces only two MT errors, associated with \textit{JavaImportOperator} and \textit{AddFieldOperator}. For \gptfive{}, the error distribution is more concentrated: \textit{LocalVariableDeclarationOperator} is the main source of MT errors, followed by \textit{AddFieldOperator} and \textit{TopLevelClassOperator}. Notably, \textit{CommentsOperator} does not produce any MT error for \gptfive{} in this run. This suggests that \gptfive{} is less affected by comment-only changes in this run, whereas structural edits involving local declarations, added fields, or top-level rearrangements remain more challenging.

\begin{table}[t]
\centering
\caption{Metamorphic-testing errors by operator (\textit{AddFieldOperator} (AF), \textit{CommentsOperator} (CO), \textit{InnerClassOperator} (IC), \textit{JavaImportOperator} (JI), \textit{LocalVariableDeclarationOperator} (LVD), and \textit{TopLevelClassOperator} (TLC)). Counts are descriptive and not normalized by operator frequency. CE = Compilation Error; BC = Behavioral Change.}
\label{tab:mt-operators}
\small
\begin{tabular}{llrrrrrr}
\hline
\textbf{Model} & \textbf{Subset} & \textbf{AF} & \textbf{CO} & \textbf{IC} & \textbf{JI} & \textbf{LVD} & \textbf{TLC} \\
\hline
\gptoss{}  & BC & 1 & 0 & 0 & 1 & 0 & 0 \\
\gptoss{}  & CE & 3 & 8 & 11 & 9 & 8 & 4 \\
\hline
\gptfive{} & BC & 1 & 0 & 0 & 1 & 2 & 0 \\
\gptfive{} & CE & 2 & 0 & 1 & 1 & 3 & 2 \\
\hline
\end{tabular}
\end{table}

Taken together, these findings indicate that both models are robust to the specific metamorphic perturbations evaluated in this study. Performance remains similar after transformation, and several difficult CE cases remain difficult in both settings. This suggests that the applied perturbations did not substantially disrupt the signals used by the models in this benchmark.
\revisionTwo{However, these results should not be interpreted as strong evidence against memorization, data contamination, or higher-order clone recognition. The transformations are mostly behaviorally inert and may preserve enough semantic, token-level, or structural similarity for a model that has seen the original or related examples to recognize them. Thus, metamorphic testing provides evidence of stability under the tested perturbations, but not evidence that the benchmark is free from contamination.
}

\begin{tcolorbox}[
    colback=gray!8,
    colframe=black!70,
    title={RQ$_{3}$ Answer},
    fonttitle=\bfseries,
    sharp corners,
    boxrule=0.6pt
]
\noindent
The performance of both models is only marginally affected by the tested semantics-preserving metamorphic transformations. \gptfive{} remains almost unchanged under metamorphic testing, while \gptoss{} also shows stable overall performance, despite a small shift between BC and CE results. Moreover, many hard CE cases remain difficult after transformation, indicating that the applied perturbations do not substantially change which compilation-error cases are challenging. \revisionTwo{However, these results should be interpreted as robustness evidence under the specific transformations evaluated in this study, not as evidence against memorization, data contamination, or higher-order clone recognition.}
\end{tcolorbox}

\section{Discussion}
\label{sec:discussion}

Next we discuss the results of our work, with emphasis on the scope and limitations of the observed evidence.

\subsection{Accuracy}

\revision{
When comparing \gptoss{} (Figure~\ref{fig:results-gpt-oss-attempts}) and \gptfive{} (Figure~\ref{fig:gpt-results}), \gptfive{} shows higher accuracy and agreement overall. At $k{=}1$, \gptfive{} reaches Acc@ $0.938$, compared with $0.805$ for \gptoss{}; at $k{=}5$, the values increase to $0.973$ and $0.929$, respectively. The main difference is in compilation-error cases, where \gptfive{} consistently obtains higher CE@ values. For behavioral-change cases, the gap is smaller: \gptfive{} starts higher, but \gptoss{} reaches full BC coverage by $k{=}3$. Agreement metrics show the same trend. \gptfive{} maintains higher tar@ and cons@ across repeated attempts, indicating more stable predictions, whereas \gptoss{} benefits more from repeated sampling but varies more across runs. Overall, \gptfive{} is more accurate and stable in this benchmark, especially for compilation-error detection, while \gptoss{} remains competitive on behavioral-change cases when multiple attempts are considered.

Both \gptoss{} and \gptfive{} correctly identify the behavioral change shown in Figure~\ref{fig:exampleOWC} (ID~14) in all five runs when applying the \textit{Push Down Method} refactoring using \eclipse{} JDT. Next, we show an example output produced by \gptfive{}. Besides returning the correct label, the model explains why the behavior changes and generates a JUnit test that compiles against both versions and exposes the expected difference: it passes before the refactoring and fails after it. This additional evidence may help developers inspect why the transformation is behavior-changing.
}

\begin{modeloutput}
\noindent\textbf{\texttt{verdict:}} NO - BEHAVIOR CHANGE

\vspace{0.4em}
\noindent\textbf{\texttt{explanation:}} The refactoring moved method \texttt{m()} from class \texttt{B} to class \texttt{C}. In the original program, \texttt{m()} calls \texttt{super.k()}, which invokes \texttt{A.k()} and returns \texttt{10}. In the refactored program, \texttt{super} refers to \texttt{B}, so \texttt{m()} calls \texttt{B.k()} and returns \texttt{20}, changing the observable output.

\vspace{0.4em}
\noindent\textbf{\texttt{junit\_test:}}

\vspace{0.2em}
\noindent
{\normalfont\ttfamily\small
\textcolor{blue}{import static} org.junit.Assert.assertEquals;\\
\textcolor{blue}{import} org.junit.Test;\\[0.3em]
\textcolor{blue}{public class} RefactoringBehaviorTest \{\\
\hspace*{1em}@Test\\
\hspace*{1em}\textcolor{blue}{public void} testMBehavior() \{\\
\hspace*{2em}assertEquals(10, \textcolor{blue}{new} C().m());\\
\hspace*{1em}\}\\
\}
}
\end{modeloutput}

The \textit{Inline Variable} refactoring performed with \netbeans{}, shown in Figure~\ref{fig:exampleOWC3} (ID~104), illustrates a compilation-error case. The refactored program does not compile because the conditional expression \texttt{flag ? 1 : 2} has primitive type \texttt{int}; therefore, the method invocation \texttt{.byteValue()} is invalid, since primitive values do not have members. \gptoss{} correctly reports this compilation error in all five runs and provides an explanation consistent with the relevant Java typing rule.

\begin{modeloutput}
\noindent\textbf{\texttt{verdict:}} NO - COMPILATION ERROR 

\vspace{0.4em}
\noindent\textbf{\texttt{explanation:}} The expression \texttt{(flag ? 1 : 2)} has type \texttt{int}. In Java, primitives do not have methods, so calling \texttt{.byteValue()} on an \texttt{int} is a compile-time error. The original code relies on autoboxing to \texttt{Integer}, which is not applied in the refactored form.
\end{modeloutput}

In contrast, \gptfive{} fails to identify this refactoring bug in all five runs. It consistently classifies the transformation as behavior-preserving and incorrectly assumes that the conditional expression has type \texttt{Integer} due to boxing. Its explanation is reproduced below.

\begin{modeloutput}
\noindent\textbf{\texttt{verdict:}} YES 

\vspace{0.4em}
\noindent\textbf{\texttt{explanation:}} The refactored code compiles: the conditional expression \texttt{(flag ? 1 : 2)} has type \texttt{Integer} due to boxing, so calling \texttt{.byteValue()} is valid. Behavior is unchanged because the local variable \texttt{iii} was not used for anything other than immediately invoking \texttt{byteValue()}, and both versions evaluate the same conditional and method call with the same result and side effects.
\end{modeloutput}

\begin{figure*}[htbp]
  \centering
  \begin{subfigure}[t]{0.47\textwidth}
    \caption{Original program}
    \label{lst:orig21}
\begin{lstlisting}[style=refjava]
public class A {
  private void compIndex(boolean flag) {
    Integer iii = flag ? 1 : 2;
    iii.byteValue();
  }
}
\end{lstlisting}
  \end{subfigure}
  \hfill
  \begin{subfigure}[t]{0.47\textwidth}
    \caption{Refactored program}
    \label{lst:refac23}
\begin{lstlisting}[style=refjava]
public class A {
  private void compIndex(boolean flag) {
    (flag ? 1 : 2).byteValue();
  }
}
\end{lstlisting}
  \end{subfigure}
  \caption{Applying \textit{Inline Variable} to \texttt{iii} using \netbeans{} introduces a compilation error.}
  \label{fig:exampleOWC3}
\end{figure*}

\subsection{Temperature}
\label{sec:discussion-temp}

Temperature is a key hyperparameter in foundation models that controls the randomness of generated outputs~\cite{temperature}. Lower values make the model more deterministic, favoring high-probability outputs and reducing variation, whereas higher values increase randomness and may produce more diverse but potentially less stable responses. Evaluating different temperature settings is therefore important to understand how the model balances sensitivity and variability in a task where consistency is important.

\revision{
To assess the sensitivity of \gptoss{} to decoding variability, we evaluated temperatures ranging from $0.0$ to $1.0$ in increments of $0.1$. Figure~\ref{fig:temperature-accuracy} summarizes the accuracy results for the full benchmark as well as for the two subsets considered in our study: behavioral-change (BC) and compilation-error (CE) instances.
Overall accuracy remains relatively stable across temperatures, ranging from $0.757$ at $t{=}0.2$ to $0.805$ at $t{=}0.5$. This limited variation suggests that \gptoss{} is relatively insensitive to temperature changes on the full benchmark. The best overall result is obtained at $t{=}0.5$, where the model correctly identifies 182 out of \bugs{} instances.

A more fine-grained analysis, however, reveals different sensitivity patterns for BC and CE cases. BC accuracy varies from $0.780$ at $t{=}0.8$ to $0.927$ at $t{=}0.1$ and $t{=}0.5$, corresponding to a spread of $14.6$\%{}. CE accuracy is more stable, ranging from $0.724$ at $t{=}0.2$ to $0.784$ at $t{=}0.8$. Thus, although the overall results appear relatively stable, BC cases are more sensitive to decoding configuration than CE cases.
The best BC performance is achieved at $t{=}0.1$ and $t{=}0.5$, where the model correctly identifies 38 out of \bugsBC{} instances. By contrast, the best CE performance is obtained at $t{=}0.8$, where \gptoss{} correctly identifies 145 out of \bugsCE{} instances. However, $t{=}0.8$ also yields the weakest BC performance, with only 32 out of \bugsBC{} BC cases correctly identified. This suggests a trade-off between the two subsets: the temperature that maximizes CE accuracy is not the best choice for BC detection. In contrast, $t{=}0.5$ provides the highest overall accuracy while maintaining the best BC accuracy and near-best CE accuracy.

The error patterns help explain this difference. For BC instances, the dominant failure mode is not only an incorrect semantic classification, but also the inability to construct an effective supporting test. Across temperatures, BC errors frequently arise because the generated test does not compile or because it compiles but does not expose the behavioral difference. Across all temperatures, BC errors include cases in which the model states that behavior was preserved, cases in which the generated test does not compile, and cases in which the generated test does not expose a behavioral change. This suggests that, for BC cases, the bottleneck involves both behavioral assessment and the construction of a valid and discriminative test oracle.
For CE instances, the dominant failure mode is an incorrect verdict, with the model repeatedly stating that the program changed behavior or preserved behavior when the correct label is compilation error. Thus, CE mistakes stem less from inadequate test generation and more from repeatedly treating uncompilable transformations as if they were semantically analyzable. This pattern is consistent across temperatures and helps explain the limited variation in CE accuracy: changing the decoding randomness does not eliminate the same underlying classification errors.

The distribution of error IDs suggests that only a limited subset of instances is persistently difficult for \gptoss{} across decoding configurations. In the BC subset, no instance is misclassified at all temperatures, indicating that every behavioral-change case can be correctly identified under at least one temperature setting. Still, some BC instances are more challenging than others, most notably ID~281, which is misclassified under most temperature settings. In contrast, the CE subset exhibits a small but important core of persistent failures. Some CE instances are misclassified at every temperature, such as IDs~94 and~298. Several additional CE instances recur in 10 out of 11 temperatures, such as IDs~15 and~105. This indicates that some uncompilable transformations repeatedly lead the model to an incorrect classification path, regardless of decoding randomness.

% IDs de CEs: 55, 94, 298, 345, 349, 396, 398
At the same time, considering the union of all temperatures, \gptoss{} correctly handles 219 out of \bugs{} instances at least once. This includes all \bugsBC{} BC cases and 178 out of \bugsCE{} CE cases. Therefore, most failures are temperature-sensitive rather than universally hard. The remaining unsolved cases are concentrated entirely in the CE subset, reinforcing that the most persistent limitation is the model's difficulty in recognizing certain compilation errors.

Taken together, the results reveal a trade-off across subsets. Lower temperatures around $t{=}0.1$ are favorable for BC detection, whereas $t{=}0.8$ provides the highest CE accuracy but harms BC performance. When selecting a single decoding configuration for \gptoss{}, $t{=}0.5$ offers the best observed balance between accuracy and stability: it achieves the highest overall accuracy, matches the best BC accuracy, and remains close to the best CE accuracy.

\begin{figure}[htbp]
    \centering
    \begin{tikzpicture}
        \begin{axis}[
            width=0.85\textwidth,
            height=0.55\textwidth,
            xlabel={Temperature},
            ylabel={Accuracy (\%)},
            xmin=0, xmax=1,
            ymin=72, ymax=94,
            xtick={0,0.1,0.2,0.3,0.4,0.5,0.6,0.7,0.8,0.9,1},
            ytick={72,74,76,78,80,82,84,86,88,90,92,94},
            grid=major,
            legend style={at={(1.02,0.5)}, anchor=west},
            legend cell align={left}
        ]

        \addplot+[thick, mark=*, color=blue] coordinates {
            (0.0,76.1) (0.1,77.9) (0.2,75.7) (0.3,77.0) (0.4,79.2) (0.5,80.5) (0.6,79.2) (0.7,80.1) (0.8,78.3) (0.9,78.3) (1.0,77.9)
        };
        \addlegendentry{Overall}

        \addplot+[thick, mark=square*, color=red] coordinates {
            (0.0,87.8) (0.1,92.7) (0.2,90.2) (0.3,90.2) (0.4,87.8) (0.5,92.7) (0.6,90.2) (0.7,90.2) (0.8,78.0) (0.9,85.4) (1.0,82.9)
        };
        \addlegendentry{BC}

        \addplot+[thick, mark=triangle*, color=teal] coordinates {
            (0.0,73.5) (0.1,74.6) (0.2,72.4) (0.3,74.1) (0.4,77.3) (0.5,77.8) (0.6,76.8) (0.7,77.8) (0.8,78.4) (0.9,76.8) (1.0,76.8)
        };
        \addlegendentry{CE}

        \end{axis}
    \end{tikzpicture}
    \caption{Accuracy across decoding temperatures for \gptoss{} on the full benchmark, behavioral-change cases (BC), and compilation-error cases (CE).}
    \label{fig:temperature-accuracy}
\end{figure}
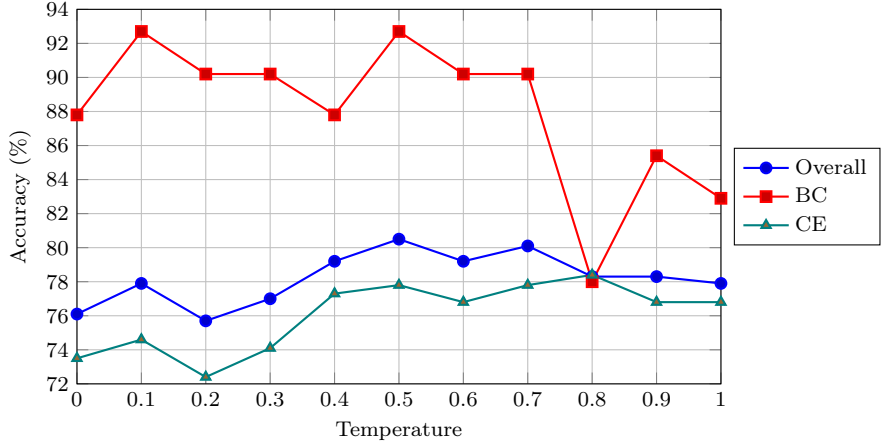

We also performed an additional \gptfive{} run with \(T=0\), keeping all other parameters used before unchanged. In this deterministic decoding setting, \gptfive{} obtained worse results than under the default-temperature configuration: 88.5\% overall accuracy, with 58.5\% accuracy on behavioral-change cases and 95.1\% on compilation-error cases. Thus, compilation-error performance remained similar, whereas behavioral-change detection dropped substantially. In the behavioral-change subset, \gptfive{} incorrectly classified 15 cases as behavior-preserving and 2 cases as introducing compilation errors.
\revisionTwo{
This result suggests that behavioral-change detection is more sensitive to decoding configuration than compilation-error detection in our benchmark. We do not interpret this difference as evidence about the model's internal reasoning process. Rather, it indicates that the reported BC performance depends on the adopted inference configuration. Consequently, higher BC accuracy obtained under the default, stochastic decoding configuration should not be interpreted as evidence that the model provides equally reliable deterministic judgments under greedy decoding. This sensitivity weakens the reliability guarantees that can be associated with BC predictions under stochastic decoding, especially in settings that require reproducible single-run decisions.
}

}

\subsection{Evaluation on Correct Refactorings (True Positives)}
\label{sec:true-positives}

Beyond detecting behavioral changes, it is important to assess whether models correctly recognize behavior-preserving transformations. For this purpose, we used a subset of the dataset from prior work~\cite{pedro-sbes-2025} covering 10 refactoring types. Each transformation applies a single refactoring to a small program generated by \jdolly{}~\cite{Soares-TSE-2013}, an automatic program generator. We evaluated 50 transformations spanning \textit{Add Parameter}, \textit{Encapsulate Field}, \textit{Move Method}, \textit{Pull Up Field}, \textit{Pull Up Method}, \textit{Push Down Field}, \textit{Push Down Method}, \textit{Rename Field}, \textit{Rename Method}, and \textit{Rename Class}. All transformations were independently validated as behavior-preserving by \saferefactor{}~\cite{saferefactor-ieee,Soares-TSE-2013,mongiovi-scaling-skips-icsme14} and by successfully compiling both the original and refactored programs.

\revision{
On the 50 true-positive refactoring instances, \gptoss{} correctly classified 43 cases, yielding an accuracy of 86\%. The remaining 7 errors were concentrated in \textit{Add Parameter}, \textit{Move Method}, \textit{Pull Up Field}, and \textit{Push Down Field} types, and mostly reflected situations in which the model inferred a behavioral difference from changes in dynamic dispatch or overload resolution, or incorrectly concluded that the refactored program did not compile. In contrast to its earlier behavior, \gptoss{} handled the other refactoring types consistently well. 

In this complementary true-positive study, \gptfive{} was evaluated with reasoning effort set to medium, whereas the main benchmark used reasoning effort none. \gptfive{} correctly classified 48 out of 50 transformations, yielding an accuracy of 96\%. The two errors occurred for \textit{Push Down Field} and \textit{Rename Field}. In both cases, the model incorrectly claimed that the transformation changed behavior and generated a JUnit test that relied on reflection to access a field, thereby exposing a difference that falls outside the intended notion of equivalence for this benchmark. In terms of efficiency, the 50 analyses consumed a total of 77,370 tokens, including 30,239 reasoning tokens, at a total cost of USD~0.72. On average, each instance used 1,547.4 total tokens and 604.8 reasoning tokens. The total runtime was approximately 1,258.7\,s, corresponding to a mean latency of 25.2\,s per instance.

These results indicate that \gptfive{} and \gptoss{} can recognize behavior-preserving refactorings with high accuracy in this additional analysis, but still benefit from a clearly specified equivalence criterion. In particular, the false positives suggest that the model may overapproximate externally visible behavior by considering reflective access to members that are not part of the intended public interface. This reinforces the importance of precise prompting when evaluating behavioral equivalence under refactoring.
}

\subsection{Other \revisionTwo{Open-Weight} Models}
\label{sec:discussion-other-open-models}

\begin{figure}[htbp]
    \centering
    \begin{tikzpicture}
        \begin{axis}[
            width=0.85\textwidth,
            height=0.55\textwidth,
            xlabel={},
            ylabel={Accuracy (\%)},
            xmin=0.5, xmax=5.5,
            ymin=0, ymax=100,
            xtick={1,2,3,4,5},
            xticklabels={\llama{}, \phimodel{}, \gptoss{}, \gemmaNew{}, \qwen{}},
            x tick label style={rotate=30, anchor=east},
            grid=major,
            legend style={at={(1.02,0.5)}, anchor=west},
            legend cell align={left}
        ]

        \addplot+[only marks, mark=*, color=blue, mark size=3pt] coordinates {
            (1,0.0)
            (2,27.9)
            (3,79.6)
            (4,96.5)
            (5,75.7)
        };
        \addlegendentry{Overall}

        \addplot+[only marks, mark=square*, color=red, mark size=3pt] coordinates {
            (1,0.0)
            (2,14.6)
            (3,87.8)
            (4,97.6)
            (5,90.2)
        };
        \addlegendentry{BC}

        \addplot+[only marks, mark=triangle*, color=teal, mark size=3pt] coordinates {
            (1,0.0)
            (2,30.8)
            (3,77.8)
            (4,96.2)
            (5,72.4)
        };
        \addlegendentry{CE}

        \end{axis}
    \end{tikzpicture}
    \caption{Overall, behavioral-change (BC), and compilation-error (CE) accuracy of the evaluated \revisionTwo{open-weight} models at temperature $0.5$.}
    \label{fig:other-models-accuracy}
\end{figure}
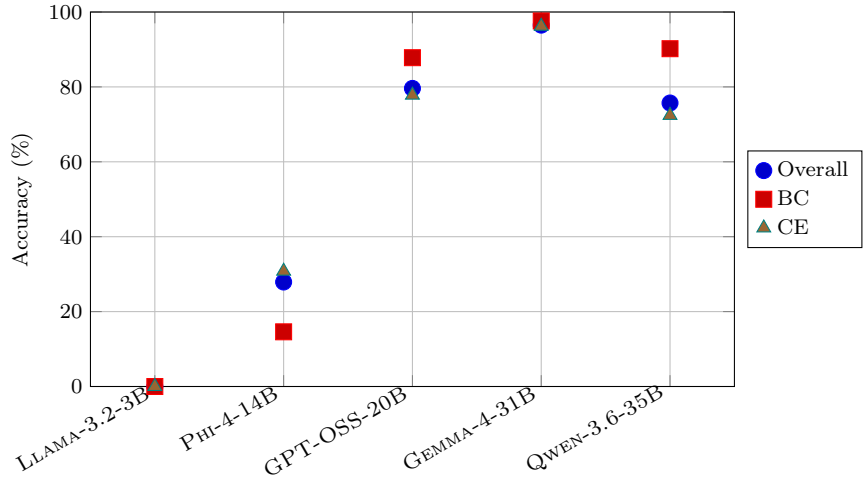

We also evaluated other \revisionTwo{open-weight} models using the same dataset, prompt, and methodology adopted for \gptoss{} in Section~\ref{sec:methodology}. More specifically, we ran \llama{}, \phimodel{}, \gemmaNew{}, and \qwen{} via Ollama, using each model's default parameters with a fixed temperature of \revision{$0.5$}. Figure~\ref{fig:other-models-accuracy} summarizes the results in three settings: overall accuracy, accuracy on behavioral changes (BC), and accuracy on compilation errors (CE). 

\revision{
The results show variation among the evaluated \revisionTwo{open-weight} models. \llama{} did not solve any benchmark instance, obtaining $0.0\%$ overall accuracy. \revisionTwo{We include this result only as a lower-bound reference for model capacity under our prompt and evaluation protocol, not as a meaningful competitive baseline or as evidence about small models in general.} Its failures included incorrect semantic verdicts, invalid top-level labels, malformed outputs marked as \texttt{PARSE\_ERROR}, and non-compiling generated tests. Thus, under our prompt and evaluation protocol, \llama{} did not reliably follow the required output format or distinguish behavioral changes from compilation errors.
\phimodel{} also achieved limited performance, with $27.9\%$ overall accuracy. It was particularly weak on behavior-change instances, correctly classifying only $14.6\%$ of them, and performed slightly better on compilation-error instances, with $30.8\%$ correctly classified. Some outputs were also marked as \texttt{PARSE\_ERROR}, indicating malformed JSON or outputs that could not be reliably parsed.
\qwen{} performed better, correctly classifying $75.7\%$ of the cases. Its performance was stronger on behavior-change instances ($90.2\%$) than on compilation-error instances ($72.4\%$). 
\gptoss{} achieved a similar overall result (\gpttossAcc{}), but with a more balanced distribution across subsets.

\gemmaNew{} achieved the strongest results among the evaluated \revisionTwo{open-weight} models, with $96.5\%$ overall accuracy. It performed consistently well on both subsets, correctly classifying $97.6\%$ of BC instances and $96.2\%$ of CE instances. These results indicate that \gemmaNew{} was effective at distinguishing behavior-change cases from compilation-error cases while also following the required output format. Under the single-run configuration used for this comparison, \gemmaNew{} also outperformed the first-run results of \gptfive{} presented in Section~\ref{sec:results}.
}

\subsection{Other Proprietary Models}
\label{sec:discussion-claude-gemini}

\revision{
We evaluated two additional proprietary large language models using the same dataset, prompt, and methodology described in Section~\ref{sec:methodology}: \gemini{} and \claude{}. \gemini{} was executed via API with default parameters, whereas \claude{} was evaluated manually through the web interface, also with default parameters. All analyses were conducted in April~2026.

Table~\ref{tab:model-comparison} summarizes the first-run results for all evaluated models, including \gptoss{} and \gptfive{} for comparison. Overall, \gemini{} achieved the highest observed performance, with 225 correct classifications out of 226 cases, corresponding to an overall accuracy of \geminiAcc{}. 
This near-perfect single-run result should not be interpreted as evidence against prior training-data exposure, given that the benchmark includes publicly reported bugs dating back to 2005 and builds on the dataset of Wang et al.~\cite{wang2024empiricalstudyrefactoringengine}. Because \gemini{} made only one error, we treat it primarily as a descriptive reference and exclude it from the formal paired significance analysis, as discussed in Section~\ref{sec:discussion-statistics}.
It was followed by \claude{} with \claudeAcc{}, \gptfive{} with \gptfiveAcc{}, and \gptoss{} with \gpttossAcc{}{}. These results indicate a clear performance gap between the strongest proprietary model in this evaluation and the remaining models, particularly the smaller open-weight model considered in our study.

The per-category results provide a more detailed view of these differences. For behavioral-change (BC) cases, \gemini{} achieved the highest accuracy, correctly handling 40 out of 41 instances. The other three models obtained the same BC accuracy. However, their failures differed. \gemini{} missed only ID~50, where the generated test did not compile. \gptfive{} also failed on ID~50, but by predicting a compilation error instead of a behavioral change, and additionally failed on IDs~112 and~281. \claude{} failed on IDs~20, 75, and~302 by stating that behavior was preserved. \gptoss{} failed on IDs~72, 97, and~263 because the generated tests did not expose a behavioral change.
For compilation-error (CE) instances, the differences between models were more pronounced. \gemini{} correctly classified all 185 CE cases, achieving perfect CE accuracy. \claude{} made 9 CE errors, while \gptfive{} made 11 CE errors. In contrast, \gptoss{} made 41 CE errors. Although \gptoss{} matched \gptfive{} and \claude{} on BC accuracy, most of its overall performance gap came from its substantially weaker ability to recognize compilation errors.

These results suggest that BC cases remain challenging because correctness depends not only on identifying that behavior changed, but also on generating a valid test that compiles and exposes the difference between the original and resulting programs. CE cases pose a different challenge: the model must recognize that the resulting program is uncompilable rather than interpreting the transformation as behavior-preserving or behavior-changing. Overall, the strongest proprietary models achieved high accuracy on both categories, with \gemini{} obtaining the best observed result in this evaluation.
}

\begin{table}[t]
\centering
\caption{Results for the four main models in the first run.}
\label{tab:model-comparison}
\small
\resizebox{\columnwidth}{!}{%
\begin{tabular}{lccc}
\hline
\textbf{Model} & \textbf{Overall Accuracy} & \textbf{BC Accuracy} & \textbf{CE Accuracy} \\
\hline
\gptoss{}  & 182/\bugs{} (0.805) & 38/\bugsBC{} (0.927) & 144/\bugsCE{} (0.778) \\
\gptfive{} & 212/\bugs{} (0.938) & 38/\bugsBC{} (0.927) & 174/\bugsCE{} (0.941) \\
\gemini{}  & 225/\bugs{} (0.996) & 40/\bugsBC{} (0.976) & 185/\bugsCE{} (1.000) \\
\claude{}  & 214/\bugs{} (0.947) & 38/\bugsBC{} (0.927) & 176/\bugsCE{} (0.951) \\
\hline
\end{tabular}%
}
\end{table}

\subsection{Mixture of Experts}
\label{sec:discussion-mixture}

\begin{figure}[t]
 \centering
 \resizebox{0.95\columnwidth}{!}{%
 \begin{tikzpicture}[font=\small\sffamily]
 \definecolor{vennBlue}{RGB}{95,165,190}
 \definecolor{vennOrange}{RGB}{242,171,76}
 \definecolor{vennGreen}{RGB}{170,200,160}
 \definecolor{vennPurple}{RGB}{190,160,210}
 \definecolor{vennStroke}{RGB}{60,60,60}
 \tikzset{
 vennOutline/.style={draw=vennStroke, line width=0.9pt},
 vennLabel/.style={font=\small\sffamily},
 vennCount/.style={font=\bfseries\sffamily, text=black}
 }

 \fill[vennPurple, opacity=0.30] (-1.6, 0.3) ellipse (2.6 and 1.4);
 \fill[vennOrange, opacity=0.30] ( 1.6, 0.3) ellipse (2.6 and 1.4);
 \fill[vennGreen, opacity=0.30] ( 0.0, 1.2) ellipse (2.2 and 2.4);
 \fill[vennBlue, opacity=0.30] ( 0.0, -1.2) ellipse (2.2 and 1.8);

 \draw[vennOutline] (-1.6, 0.3) ellipse (2.6 and 1.4);
 \draw[vennOutline] ( 1.6, 0.3) ellipse (2.6 and 1.4);
 \draw[vennOutline] ( 0.0, 1.2) ellipse (2.2 and 2.4);
 \draw[vennOutline] ( 0.0, -1.2) ellipse (2.2 and 1.8);

 \node[vennLabel] at (-5.0, 1.3) {\claude{}};
 \node[vennLabel] at ( 5.0, 0.3) {\gptfive{}};
 \node[vennLabel] at ( 3.0, 3.3) {\gemini{}};
 \node[vennLabel] at ( 0.0, -3.3) {\gptoss{}};

 \node[vennCount] at ( 0.00, 0.15) {163};
 \node[vennCount] at ( 0.00, 0.90) {39};
 \node[vennCount] at (-1.20, -0.05) {10};
 \node[vennCount] at ( 1.20, -0.05) {6};
 \node[vennCount] at ( 1.30, 1.05) {4};
 \node[vennCount] at (-1.30, 1.05) {1};
 \node[vennCount] at (-1.8, -0.80) {1};
 \node[vennCount] at ( 0.00, -1.0) {2};

 \begin{scope}[shift={(0,-4.5)}] 
 \fill[vennPurple, opacity=0.30, draw=vennStroke, line width=0.5pt]
 (-3.9, 0.4) rectangle (-3.5, 0.62);
 \node[anchor=west, vennLabel] at (-3.5, 0.51) {\claude{}};

 \fill[vennOrange, opacity=0.30, draw=vennStroke, line width=0.5pt]
 ( 0.5, 0.4) rectangle ( 0.9, 0.62);
 \node[anchor=west, vennLabel] at ( 1.0, 0.51) {\gptfive{}};

 \fill[vennGreen, opacity=0.30, draw=vennStroke, line width=0.5pt]
 (-3.9, 0.0) rectangle (-3.5, 0.22);
 \node[anchor=west, vennLabel] at (-3.5, 0.11) {\gemini{}};

 \fill[vennBlue, opacity=0.30, draw=vennStroke, line width=0.5pt]
 ( 0.5, 0.0) rectangle ( 0.9, 0.22);
 \node[anchor=west, vennLabel] at ( 1.0, 0.11) {\gptoss{}};
 \end{scope}
 \end{tikzpicture}%
 }
 \caption{Diagram of bugs correctly detected by each model. Numbers indicate the number of bugs in each overlap region.}
 \label{fig:venn-moe-4models}
\end{figure}
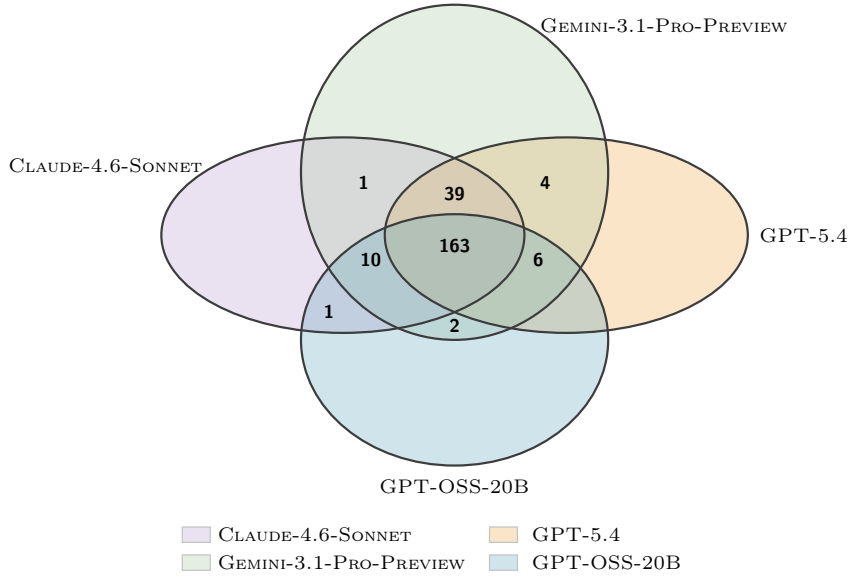

\revision{
We also evaluated a simple mixture-of-experts (MoE) strategy over the four models considered in this analysis: \gptoss{}, \gptfive{}, \gemini{}, and \claude{}. In this setting, an instance is considered correctly solved if at least one model succeeds; that is, we combine model outputs using a logical OR. Since the results reported here rely on the first attempt of each model, this MoE analysis should be interpreted as a complementary analysis of cross-model agreement and residual-error diversity rather than as a repeated-sampling strategy over multiple runs.

Figure~\ref{fig:venn-moe-4models} summarizes the overlap among the bugs correctly detected by each model. The four models jointly solved 163 instances, forming a large common core of cases that were consistently handled across model families. In addition, 39 instances were solved by the three strongest proprietary models, namely \gemini{}, \gptfive{}, and \claude{}, but missed by \gptoss{}. This region captures most of the gap between \gptoss{} and the proprietary models. Smaller regions reveal complementary behavior among the remaining systems: 10 instances were solved by \gemini{}, \claude{}, and \gptoss{} but missed by \gptfive{}, while 6 instances were solved by \gemini{}, \gptfive{}, and \gptoss{} but missed by \claude{}. The remaining non-empty regions were smaller: 4 instances were solved only by \gemini{} and \gptfive{}, 2 only by \gemini{} and \gptoss{}, 1 only by \gemini{} and \claude{}, and 1 only by \claude{} and \gptoss{}.

A simple OR-based MoE is therefore effective in this benchmark, but its interpretation should be limited to an upper-bound complementarity analysis: it assumes that an external mechanism can identify or select a correct answer whenever at least one model produces one. Nevertheless, the result is useful because it shows that combining model outputs can recover residual errors left by individual systems. This finding motivates future work on cost-aware MoE strategies, such as cascaded evaluation pipelines in which a cheaper or faster model is queried first and stronger models are invoked only for uncertain or unresolved cases.
}

\subsection{Costs}

Cost is an important consideration when deploying these models at scale. As with the execution-time analysis, our goal is not to provide a definitive or exact cost benchmark. Costs depend on several factors, including hardware availability, cloud provider pricing, API pricing, batching strategy, implementation overhead, and changes in model-serving infrastructure. Therefore, the values reported here should be interpreted as indicative estimates under the conditions of our evaluation and current pricing. We also expect these costs to decrease over time as inference infrastructure improves, hardware becomes more efficient, and providers adjust pricing.

\revision{
Although we run \gptoss{} locally on a Mac Mini M4 Pro, estimating an equivalent cloud rental cost still provides useful perspective on resource efficiency. Based on the listed rental price of a Mac Mini cloud instance, EUR~0.22 per hour\footnote{\url{https://www.scaleway.com/en/mac-mini-m4/}}, and on the recorded end-to-end runtime of the first \gptoss{} run over the \bugs{} instances, the estimated infrastructure cost is approximately EUR~0.31 per complete run. 
In this run, the mean per-call latency was 22.27\,s, the median was 13.61\,s, the minimum was 5.52\,s, the maximum was 512.67\,s, and the total runtime was 1.40\,h. These values account only for infrastructure rental time and do not include engineering time, storage, network transfer, or operational overhead. This indicates that \gptoss{} can be evaluated at low monetary cost when suitable local or low-cost cloud hardware is available, although its per-instance latency is higher than that of \gptfive{}.

For \gptfive{}, the average cost was approximately \${}0.79 per complete benchmark run. Across these runs, \gptfive{} consumed approximately 180K tokens per run over the \bugs{} benchmark, comprising about 153K input tokens and 27K output tokens per attempt. The input-token cost is largely fixed by the benchmark prompts, while the small variation in total cost across runs is mainly attributable to differences in generated output length.

These numbers highlight a deployment-relevant trade-off between the two main configurations considered in this cost discussion. Compared with the local \gptoss{} setup, \gptfive{} offers higher observed effectiveness and lower per-call latency, but at a higher monetary cost per complete run. At the same time, the API-based deployment reduces operational overhead because it does not require provisioning or maintaining local inference hardware. This trade-off is especially relevant in scenarios requiring repeated executions, such as pass@k analyses, temperature-sensitivity studies, metamorphic-testing sensitivity analyses, or large-scale benchmarks.
}
\revisionTwo{
Table~\ref{tab:cost-summary} summarizes the estimated monetary cost per complete benchmark run. Section~\ref{sec:execution-time} separately reports detailed latency statistics for \gptoss{}, which help contextualize the infrastructure-time component of the cost estimate reported here.
}

\begin{table}[t]
\centering
\caption{\revisionTwo{Estimated cost for the first benchmark run of each model.}}
\label{tab:cost-summary}
\small
\begin{tabular}{llrr}
\hline
\textbf{Model} & \textbf{Cost basis} & \textbf{Resource/run} & \textbf{Cost/run} \\
\hline
\gptoss{}  & Infrastructure rental & 1.40\,h & EUR~0.31 \\
\gptfive{} & API token usage & 153K input + 27K output & US\$~0.79 \\
\hline
\end{tabular}
\end{table}

\subsection{Execution Time}
\label{sec:execution-time}

To provide a general sense of runtime, we report the observed latency of each model in our experiments, without aiming at a controlled benchmarking study. These measurements should therefore be interpreted as indicative rather than definitive, since they are affected by factors such as hardware configuration, API latency, implementation overhead, network conditions, and background system load.

\revision{
Table~\ref{tab:execution-time} summarizes the latency observed for each model. Among the evaluated models, \llama{} was the fastest, with a mean latency of 1.76\,s per instance and a median of 1.65\,s. This corresponds to approximately 6.6 minutes to process the full benchmark of \bugs{} transformations. However, as discussed in Section~\ref{sec:discussion-other-open-models}, this low latency came with very poor predictive performance, since \llama{} did not correctly solve any instance under our evaluation protocol.
\gptfive{} was also very fast, with a mean response time of 3.25\,s per instance and a median of 2.71\,s. It processed the full benchmark in approximately 12.2 minutes, although one outlier reached 43.75\,s. This makes \gptfive{} one of the lowest-latency models in our evaluation, combining low latency with high observed predictive performance. \phimodel{} was slower than \llama{} and \gptfive{}, but still relatively efficient, with a mean latency of 9.41\,s and a median of 7.88\,s, corresponding to approximately 35.4 minutes for the full benchmark.

\gemini{} also showed moderate latency. Its mean response time was 11.41\,s per instance and its median was 7.46\,s, corresponding to approximately 43 minutes for the full benchmark. Most responses were relatively fast, but one outlier reached 446.78\,s. Given that \gemini{} achieved the highest observed predictive performance among all evaluated models, this result indicates a favorable observed accuracy--latency trade-off in our setting.
\gptoss{} showed higher latency than \gptfive{}, \phimodel{}, and \gemini{}, with a mean response time of 22.27\,s and a median of 13.61\,s per instance. This corresponds to approximately 1.4 hours to analyze all \bugs{} transformations. Its runtime was also more variable, with responses ranging from 5.52\,s to 512.67\,s.

The larger \revisionTwo{open-weight} models were slower. \gemmaNew{} achieved the highest observed predictive performance among the evaluated \revisionTwo{open-weight} models, and even outperformed the first-run results of \gptfive{} and \gptoss{} under the configuration used in our experiments. However, \gemmaNew{} required a mean latency of 110.73\,s per instance and a median of 69.28\,s, corresponding to approximately 7.0 hours for the full benchmark. \qwen{} also exhibited high latency, with a median response time of 82.64\,s per instance. Its runtime was highly variable, with several extreme outliers above 23,000\,s, which increased its mean latency to 611.02\,s and its estimated total runtime to approximately 38.4 hours.

\gemini{} achieved very high accuracy with moderate latency, while \gptfive{} provided high accuracy with low latency. Beyond runtime alone, these models also provide explanatory feedback about behavioral changes and compilation errors, a capability that may complement traditional refactoring implementations and may be useful in AI-assisted IDEs such as Antigravity~\cite{antigravity}, Cursor~\cite{cursor}, and Windsurf~\cite{windsurf}.
}

\begin{table}[t]
\centering
\caption{Observed execution time per instance. Total time is estimated from the sum of per-instance latencies over the \bugs{} transformations.}
\label{tab:execution-time}
\small
\resizebox{\columnwidth}{!}{%
\begin{tabular}{lrrrrr}
\hline
\textbf{Model} & \textbf{Mean} & \textbf{Median} & \textbf{Min.} & \textbf{Max.} & \textbf{Total} \\
\hline
\llama{}     & 1.76s   & 1.65s  & 1.28s  & 3.52s        & 6.6m \\
\gptfive{}   & 3.25s   & 2.71s  & 1.75s  & 43.75s       & 12.2m \\
\phimodel{}  & 9.41s   & 7.88s  & 5.53s  & 22.45s       & 35.4m \\
\gemini{}    & 11.41s  & 7.46s  & 3.57s  & 446.78s      & 43.0m \\
\gptoss{}    & 22.27s  & 13.61s & 5.52s  & 512.67s      & 1.4h \\
\gemmaNew{}  & 110.73s & 69.28s & 26.90s & 870.88s      & 7.0h \\
\qwen{}      & 611.02s & 82.64s & 24.99s & 23{,}361.43s & 38.4h \\
\hline
\end{tabular}%
}
\end{table}

\subsection{Statistical Analysis of Model Accuracy}
\label{sec:discussion-statistics}

\revision{
We evaluated four models (\gptoss{}, \gptfive{}, \gemini{}, and \claude{}) on a dataset of $N{=}\bugs{}$ bugs. For each bug, we recorded a binary outcome indicating whether the model produced the correct classification. 
Table~\ref{tab:acc-ci-226} reports the overall accuracy of each model together with 95\% Wilson confidence intervals. \gemini{} achieved very high observed accuracy, making only one error in the entire benchmark. Because this leaves almost no variability for inferential comparison, we use \gemini{} mainly as a descriptive reference in this section and focus the formal paired significance analysis on \gptoss{}, \gptfive{}, and \claude{}. This choice is also methodologically useful because these three models still exhibit a non-trivial number of errors, allowing more informative paired comparisons. Moreover, they represent three practically relevant profiles: a smaller open-weight model that can be executed locally (\gptoss{}), a strong proprietary model (\gptfive{}), and a strong alternative proprietary model from another provider (\claude{}).
\revisionTwo{
Table~\ref{tab:acc-ci-226} reports 95\% Wilson confidence intervals separately for overall, BC, and CE accuracy. As expected, the BC confidence intervals are substantially wider because this subset contains only \bugsBC{} instances, whereas the CE subset contains \bugsCE{} instances. Consequently, BC-specific estimates, particularly when stratified by refactoring type, should be interpreted cautiously.
}

\begin{table}[t]
\centering
\caption{\revisionTwo{Overall, behavioral-change (BC), and compilation-error (CE) accuracy with 95\% Wilson confidence intervals.}}
\label{tab:acc-ci-226}
\small
\resizebox{\columnwidth}{!}{%
\begin{tabular}{lccc}
\hline
\textbf{Model} & \textbf{Overall Acc. (95\% CI)} & \textbf{BC Acc. (95\% CI)} & \textbf{CE Acc. (95\% CI)} \\
\hline
\gptoss{}  & 182/\bugs{} (0.805) [0.749, 0.852] & 38/\bugsBC{} (0.927) [0.806, 0.975] & 144/\bugsCE{} (0.778) [0.713, 0.832] \\
\gptfive{} & 212/\bugs{} (0.938) [0.899, 0.963] & 38/\bugsBC{} (0.927) [0.806, 0.975] & 174/\bugsCE{} (0.941) [0.897, 0.966] \\
\claude{}  & 214/\bugs{} (0.947) [0.909, 0.969] & 38/\bugsBC{} (0.927) [0.806, 0.975] & 176/\bugsCE{} (0.951) [0.910, 0.974] \\
\gemini{}  & 225/\bugs{} (0.996) [0.975, 0.999] & 40/\bugsBC{} (0.976) [0.874, 0.996] & 185/\bugsCE{} (1.000) [0.980, 1.000] \\
\hline
\end{tabular}%
}
\end{table}

To ensure a like-for-like comparison across models, all cross-model statistical analyses in this section use the first-run outcome for each model; the additional repeated runs of \gptoss{} and \gptfive{} are not used in these comparisons.
To compare \gptoss{}, \gptfive{}, and \claude{} statistically, we used tests for paired dichotomous outcomes. 
We first applied Cochran's $Q$ test~\cite{cochran1950,conover1999} as an omnibus test across the three models. The result indicates that at least one model differs significantly from the others ($Q{=}30.60$, $p{<}0.001$). We then performed pairwise McNemar exact tests~\cite{mcnemar1947} (two-sided), with Holm--Bonferroni correction~\cite{holm1979} to control the family-wise error rate over the three pairwise contrasts.
To make the analysis fully explicit, Table~\ref{tab:mcnemar-226} reports the exact paired contingency counts for each comparison: $n_{11}$ (both models correct), $n_{10}$ ($A$ correct and $B$ wrong), $n_{01}$ ($A$ wrong and $B$ correct), and $n_{00}$ (both models wrong). We also report the accuracy difference $\Delta(A{-}B) = (n_{10}-n_{01})/\bugs{}$.

\begin{table}[t]
\centering
\caption{Pairwise two-sided exact McNemar tests with Holm correction.}
\label{tab:mcnemar-226}
\small
\resizebox{\columnwidth}{!}{%
\begin{tabular}{lccccccc}
\hline
\textbf{Pair} & \textbf{$n_{11}$} & \textbf{$n_{10}$} & \textbf{$n_{01}$} & \textbf{$n_{00}$} & \textbf{$\Delta$} & \textbf{$p_{\mathrm{exact}}$} & \textbf{$p_{\mathrm{Holm}}$} \\
\hline
\gptoss{} vs.\ \gptfive{} & 169 & 13 & 43 & 1 & $-0.133$ & $7.33{\times}10^{-5}$ & $1.47{\times}10^{-4}$ \\
\gptoss{} vs.\ \claude{}  & 174 & 8  & 40 & 4 & $-0.142$ & $3.31{\times}10^{-6}$ & $9.92{\times}10^{-6}$ \\
\gptfive{} vs.\ \claude{} & 202 & 10 & 12 & 2 & $-0.009$ & $8.32{\times}10^{-1}$ & $8.32{\times}10^{-1}$ \\
\hline
\end{tabular}%
}
\end{table}

The results show that both \gptfive{} and \claude{} outperform \gptoss{}, with absolute accuracy gains of 13.3\% and 14.2\%, respectively. In contrast, the difference between \gptfive{} and \claude{} is very small, less than 1\%, and is not statistically significant after correction. Thus, the inferential analysis supports an observed separation between \gptoss{} and the two stronger proprietary models, while \gptfive{} and \claude{} do not show a statistically significant difference on this benchmark.
For completeness, \gemini{} obtained the highest descriptive accuracy, correctly classifying 225 out of \bugs{} bugs. However, because its performance is very high, we avoid over-interpreting pairwise significance tests involving \gemini{}.
}

\subsection{Large Projects}
\label{sec:feasibility-diff}

Many codebases exceed the context window supported by current models; consequently, applying the full-source evaluation prompt from Section~\ref{sec:methodology-prompts} is impractical for such projects. \revision{As a feasibility study,} we therefore evaluate large codebases using unified diffs from real projects, following a similar idea~\cite{pedro-sbes-2025}. 
\revision{
Because this experiment involves real projects and diff-only reasoning, we treat it as a complementary feasibility study rather than as a definitive benchmark. Unlike the controlled benchmark used in the main evaluation, correctness judgments in this setting cannot always be established mechanically from the diff alone. The goal is therefore not to provide a mechanically verified benchmark over large projects, but to examine how foundation-model-based assessment behaves on realistic refactoring diffs and to identify practical difficulties that arise in this setting.
}

We adopt MetaPrompting~\cite{hou-metaprompting} to guide the model in generating or refining task prompts. Since the model does not have access to the entire source code in this setting, we do not ask it to generate a test case.

\begin{tcolorbox}[
  breakable,
  colback=gray!5,
  colframe=black!60,
  boxrule=0.5pt,
  arc=2pt,
  left=6pt,right=6pt,top=6pt,bottom=6pt
]
\footnotesize 
\noindent I used the following prompt to evaluate refactoring correctness in small programs. \\
\noindent \texttt{prompt} \\
\noindent However, due to the context window, I cannot use it to evaluate larger programs. Create a similar prompt that evaluates the refactoring correctness based on the diff between the source and refactored programs. Do not need to generate a test case when identifying a behavior change.
\end{tcolorbox}

The variable \texttt{prompt} refers to the prompt described in Section~\ref{sec:methodology-prompts}. In response, the model produced a diff-oriented prompt that targets both compilation errors and behavioral changes while retaining the same verdict schema used in the controlled benchmark. The variable \texttt{diff} includes added, removed, and minimal unchanged context lines. \revision{Because diff-only reasoning may not provide enough information to determine correctness with confidence, the model conservatively added a fourth possible verdict: \texttt{UNKNOWN}.}

\begin{tcolorbox}[
  breakable,
  colback=gray!5,
  colframe=black!60,
  boxrule=0.5pt,
  arc=2pt,
  left=6pt,right=6pt,top=6pt,bottom=6pt
]
\footnotesize
\noindent Consider the following diff between an initial program and the resulting program after applying a refactoring: \\

\noindent \texttt{<DIFF} \\
\noindent \texttt{\{diff\}} \\
\noindent \texttt{DIFF>} \\

\noindent The diff may include one or more files. Lines removed from the initial program are marked with \texttt{"-"}. Lines added in the resulting program are marked with \texttt{"+"}. Context lines are unchanged. \\

\noindent Evaluate whether the resulting program is a correct refactoring of the initial program, based only on the information available in the diff. \\

\noindent Check the following: \\

\noindent 1. Compilation correctness: \\
\noindent Determine whether the changes shown in the diff introduce an evident compilation problem in the resulting program, such as syntax errors, type errors, unresolved names, invalid imports, invalid method calls, invalid overriding, incompatible access modifiers, or linkage-related problems. \\

\noindent 2. Behavioral equivalence: \\
\noindent Determine whether the changes shown in the diff preserve the observable behavior of the common public API. Ignore methods not present in both programs; compare only methods with the same signature that are public in both versions. For those common public methods, the initial and resulting programs preserve the same observable behavior, including return values, printed output, thrown exceptions, and externally visible state changes. \\

\noindent Return ONLY valid JSON. Do not use markdown. Do not add any text before or after the JSON. \\

\noindent Use exactly this schema: \\
\noindent \texttt{\{} \\
\noindent \texttt{\ \ "verdict": "YES | NO - COMPILATION ERROR | NO - BEHAVIOR CHANGE | UNKNOWN",} \\
\noindent \texttt{\ \ "explanation": "1-4 sentences"} \\
\noindent \texttt{\}} \\

\noindent Rules: \\
\noindent - \texttt{"verdict"} must be exactly one of: \\
\noindent \texttt{"YES"} \\
\noindent \texttt{"NO - COMPILATION ERROR"} \\
\noindent \texttt{"NO - BEHAVIOR CHANGE"} \\
\noindent \texttt{"UNKNOWN"} \\
\noindent - Use \texttt{"YES"} only when the diff provides sufficient evidence that the resulting program compiles and preserves the behavior of the common public API. \\
\noindent - Use \texttt{"NO - COMPILATION ERROR"} only when the diff provides concrete evidence of a compilation problem introduced by the resulting program. \\
\noindent - Use \texttt{"NO - BEHAVIOR CHANGE"} only when the diff provides concrete evidence of a behavioral difference in a common public method, such as a changed return value, changed printed output, changed thrown exception, changed mutation of externally visible state, or changed public-visible interaction. \\
\noindent - Use \texttt{"UNKNOWN"} when the diff does not contain enough surrounding context to determine compilation correctness or behavioral equivalence with confidence. \\
\noindent - Do not invent compilation errors. \\
\noindent - Do not claim behavior change unless the diff shows a specific differing return value, output, exception, state change, or public-visible interaction. \\
\noindent - Do not assume behavior preservation merely because the change is labeled as a refactoring. \\
\noindent - Do not assume missing declarations, imports, fields, methods, or classes unless they are visible in the diff. \\
\noindent - If a symbol is added, removed, renamed, or has its type, visibility, signature, inheritance relationship, or initialization changed, consider whether the diff provides enough context to determine whether all affected uses remain valid. \\
\noindent - If a public method is added or removed, ignore that method for behavioral equivalence, but consider whether the change affects existing common public methods. \\
\noindent - Private, protected, or package-private changes are relevant only if they affect compilation or the observable behavior of common public methods. \\
\noindent - Formatting-only changes, comment-only changes, import reordering, or equivalent renamings should be considered behavior-preserving unless they create a concrete compilation or behavior issue. \\
\noindent - The \texttt{"explanation"} must briefly justify the answer with concrete evidence from the diff. \\
\noindent - Return JSON only.
\end{tcolorbox}

\revision{
We evaluated \numTransformationsReal{} refactorings applied to real-world open-source Java projects using \intellij{}~2024.1.4 (Table~\ref{tab:real-projects}). The dataset covers twelve widely used refactoring types: \textit{Extract Class}, \textit{Extract Interface}, \textit{Extract Superclass}, \textit{Inline Method}, \textit{Move Method}, \textit{Pull Up Field}, \textit{Pull Up Method}, \textit{Push Down Field}, \textit{Push Down Method}, \textit{Rename Field}, \textit{Rename Method}, and \textit{Rename Class}. Each transformation corresponds to a single refactoring instance manually applied in \intellij{}. In April~2026, we used the \gptfive{} API to adapt the full-source prompt to a diff-only setting, setting reasoning effort to medium and leaving all other configuration parameters at their default values.

\revisionTwo{
The main result of this feasibility study is that diff-only reasoning often does not provide enough context for a definitive oracle judgment. \gptfive{} returned \texttt{UNKNOWN} for 18 out of 44 transformations (40.9\%), indicating that in more than one third of the cases the model could not determine correctness from the visible patch alone. This is a practical limitation for using foundation models as standalone oracles in large projects. The \texttt{UNKNOWN} cases were mainly due to insufficient contextual information in the diffs. In several instances, public or protected members were renamed, moved, or removed, such as methods, fields, interfaces, and nested types, but the diffs did not show all project-wide references, subclasses, constructor declarations, imports, or external callers that could still depend on the old declarations. In other cases, the visible changes appeared locally consistent and no clear behavioral change was evident in common public methods, but correctness depended on unseen code, such as whether moved methods were implemented elsewhere, whether fields remained accessible across packages, whether callbacks could be null or behaviorally different, or whether all concrete subclasses still satisfied required interfaces.
}

\revisionTwo{
For the remaining cases, \gptfive{} returned \texttt{YES} for 24 transformations and \texttt{NO - BEHAVIOR CHANGE} for 2 transformations. The 24 \texttt{YES} judgments should be interpreted cautiously: they were judged by the authors as supported by the visible diff evidence, but they were not independently verified against a mechanically established ground-truth oracle. In these cases, the edited declarations, implementations, and visible call sites appeared coherent, and no concrete compilation error or behavioral difference in the common public API was evident from the diff. The two \texttt{NO - BEHAVIOR CHANGE} verdicts were not confirmed by the authors. Although the model identified potential behavioral concerns, the explanations did not provide sufficient evidence from the diff to establish a concrete behavioral difference under our adopted equivalence notion. We therefore treat these two cases as unresolved borderline cases rather than as evidence that the transformations were behavior-preserving. This conservative adjudication may undercount false negatives, because a real behavioral difference could still exist outside the visible diff context.
}
}

\begin{table}[t]
\centering
\caption{Java projects used in the large-project evaluation.}
\label{tab:real-projects}
\footnotesize
\setlength{\tabcolsep}{2.5pt}
\begin{tabular}{lp{0.32\columnwidth}rrrr}
\hline
\textbf{Project} & \textbf{Domain} & \textbf{KLOC} & \textbf{Stars (K)} & \textbf{Contr.} & \textbf{Transf.} \\
\hline
Lettuce & A scalable thread-safe Redis client for synchronous, asynchronous, and reactive usage. & 234 & 5.6 & 135 & 30 \\
Apache Gobblin & A distributed data integration framework. & 454 & 2.6 & 115 & 1 \\
Google Maps Services & A Java client for Google Maps Services. & 38 & 1.7 & 96 & 3 \\
Spring Boot & A framework to create Spring-based applications. & 674 & 77.1 & 1{,}156 & 4 \\
RefactoringMiner & A refactoring detection tool. & 127 & 0.4 & 18 & 6 \\
\hline
\end{tabular}
\end{table}

\revision{
Because the diff-only setting does not always provide enough context to mechanically establish whether a transformation preserves behavior, we manually assessed the model judgments in this analysis. Each \gptfive{} judgment was independently reviewed by two authors. The reviewers applied the same decision criterion: a judgment was considered correct only when the model's verdict was supported by evidence visible in the diff, under the behavioral-equivalence notion adopted in this study. The two initial reviewers agreed on 42 out of 44 cases. 
\revisionTwo{
The two disagreements were borderline cases involving possible API-level or visibility-related effects for which the diff did not provide enough context to establish all behavioral consequences. After discussion, the reviewers found no sufficient diff-visible evidence to confirm a behavioral change. However, because no independent whole-project oracle was available, these decisions should not be interpreted as proof of behavior preservation. They instead illustrate that author adjudication in diff-only analysis may undercount false negatives when relevant dependencies lie outside the visible patch. Therefore, we treat this analysis as a feasibility study rather than as a definitive benchmark.
}
}

While the diff-only strategy is a useful exploratory step, important challenges remain in assessing behavior preservation for real-world programs using foundation models. The approach is particularly difficult when the decisive evidence lies outside the edited regions or depends on runtime semantics that cannot always be inferred from static analysis over the diff alone. \revision{This limitation is not specific to foundation models: diff-only review can also be challenging for human reviewers when behavior preservation depends on global semantic relationships that are not visible in the transformation.} As future work, we plan to refine the prompt to make the adopted notion of behavioral equivalence more explicit and to investigate retrieval-augmented generation techniques that provide richer project context for patch analysis, building on recent work on large-scale refactoring automation~\cite{batole2025leveragingllmsidessemantic}.

\subsection{\revisionTwo{Failure Analysis}}
\label{sec:failure-analysis}

\revisionTwo{
Beyond aggregate accuracy, we analyzed the main failure modes across the five repeated attempts of \gptfive{} and \gptoss{}. Table~\ref{tab:failure-analysis} summarizes these failures by model and bug type.
}

% Preamble:
% \usepackage{multirow}

\begin{table}[t]
\centering
\caption{\revisionTwo{Failure categories across five attempts. Bug type denotes the ground-truth class: behavioral change (BC) or compilation error (CE). YES, BC, and CE count incorrect verdicts in which the model respectively classified the instance as behavior-preserving, as a behavioral change, or as a compilation error. Invalid (Inv.) test counts generated JUnit tests that did not compile, and non-detecting (Non-det.) test counts generated tests that compiled and executed but did not expose the behavioral difference. Single-run models are excluded from this analysis by design.}}
\label{tab:failure-analysis}
\footnotesize
\resizebox{\columnwidth}{!}{
\begin{tabular}{llrrrrrr}
\hline
\textbf{Model} & \textbf{Bug type} & \textbf{YES} & \textbf{BC} & \textbf{CE} & \textbf{Inv. test} & \textbf{Non-det. test} & \textbf{Total} \\
\hline
\multirow{2}{*}{\gptfive{}}
  & BC & 10 & -- & 4  & 5  & 1 & 20 \\
  & CE & 16 & 30 & -- & -- & -- & 46 \\
\hline
\multirow{2}{*}{\gptoss{}}
  & BC & 8  & --  & 0 & 7  & 8 & 23 \\
  & CE & 116 & 105 & -- & -- & -- & 221 \\
\hline
\end{tabular}
}
\end{table}

\revisionTwo{
For BC cases, the two models fail in different ways. For \gptfive{}, the dominant BC failure is an incorrect verdict: 10 out of 20 BC errors state that behavior is preserved, and 4 incorrectly classify the case as a compilation error. Test-generation failures are less frequent, with 5 non-compiling generated tests and 1 test that compiles and executes but does not expose the behavioral difference. In contrast, \gptoss{}'s BC errors are more often related to executable evidence: 7 generated tests do not compile and 8 compile and execute but do not expose the behavioral change, while 8 cases incorrectly state that behavior is preserved.

For CE cases, both models mainly fail by treating an uncompilable resulting program as semantically analyzable. \gptfive{} produces 46 CE errors across five attempts, including 30 cases classified as behavioral changes and 16 classified as behavior-preserving transformations. \gptoss{} produces more CE errors: 221 across five attempts, split between 116 behavior-preserving verdicts and 105 behavioral-change verdicts. This pattern reinforces the observation that CE detection is the main limitation of \gptoss{}, especially for transformations involving name resolution, member placement, inheritance, and compilation validity.

Overall, the failure analysis shows that model errors do not arise from a single source. They include incorrect verdicts, confusion between CE and BC labels, failure to generate executable JUnit evidence, and tests that fail to expose behavioral differences. This distinction is important because different failure categories require different mitigations: compiler integration can reduce CE misclassification, whereas stronger test-generation constraints or post-generation validation are needed for BC evidence failures.
}

\subsection{Threats to Validity}
\label{sec:threats}

Several threats to validity apply to this study~\cite{sallou2024breaking}. 

\revision{
\subsubsection{Internal Validity}
Internal-validity threats stem from prompt and context sensitivity, model non-determinism, provider-specific execution settings, ground-truth reconstruction, human adjudication, and possible data leakage. Although we used standardized prompt templates across models, small wording changes or different context representations may affect predictions. This is particularly relevant because the task requires both classification and, for behavioral-change cases, generation of an executable test oracle. We fixed prompts and evaluation scripts, ran repeated trials where feasible, and reported stability measures such as accuracy spread, cumulative coverage, tar@, and cons@. Still, we did not systematically ablate prompt wording, context order, context amount, or all decoding configurations.

Ground-truth reliability is another internal-validity threat. Not all bug reports provide self-contained examples, and reconstructing minimal reproducing programs is manual and potentially error-prone. Reproduction may depend on JDK versions, IDE versions, project configuration, dependencies, and historical tool behavior. To reduce this risk, we relied on available attachments and version information, recorded environment details, compiled all programs, validated behavior-preserving cases with \saferefactor{}, and manually rechecked ambiguous cases. We revised the dataset construction and validation procedure to make the ground truth more explicit, executable, and reproducible, and removed structurally redundant near-duplicate cases found during revalidation. Nevertheless, reconstruction remains a possible source of error.

Data leakage is a further concern. We applied metamorphic testing to all benchmark instances to assess whether intended semantics-preserving changes would substantially alter predictions. The results showed little overall degradation for \gptoss{} and \gptfive{}. However, metamorphic testing does not rule out leakage. A model may still benefit from prior exposure to related code, refactoring patterns, or abstract bug structures, and our operators cover only a subset of possible surface-level variations. At the same time, the persistence of some CE failures across repeated attempts and temperature settings is inconsistent with a simple direct-memorization explanation for those instances, although it does not rule out prior exposure. Likewise, the substantial drop in BC accuracy under greedy decoding neither establishes nor rules out memorization, but it shows that the high default-configuration BC accuracy depends on inference settings rather than reflecting a consistently reproduced outcome.

Model and infrastructure effects may also influence the results. Some models were executed through vendor APIs, others through local wrappers such as Ollama, and \claude{} was evaluated through the web interface. These settings may differ in hidden system prompts, decoding defaults, truncation behavior, rate limits, wrappers, or model revisions. Proprietary models may also change over time. Therefore, the results should be understood as a snapshot of the models and configurations available during our evaluation.

\subsubsection{Construct Validity}
Construct-validity threats arise from how we operationalize refactoring correctness, model outputs, and evaluation metrics. Our benchmark uses binary labels for compilation errors and behavioral changes, but this abstraction may hide partial correctness, uncertainty, or explanations that are technically plausible but insufficiently justified. Some model outputs contain correct-looking verdicts with weak explanations, while others identify plausible risks without enough evidence to establish a compilation error or behavioral change under our adopted equivalence notion. Disagreements may also arise from different interpretations of behavioral preservation, such as how to treat exception locations, public API changes, removed public members, or visibility changes. To mitigate this, we compiled programs, compiled and executed the model-generated tests with JUnit to check whether they actually expose the reported behavioral differences, used \saferefactor{} on behavior-preserving transformations, manually reviewed explanations, and distinguished confirmed bugs from unsupported concerns and insufficient-evidence cases. Future work should score explanation quality and explore uncertainty-aware metrics.

The main bug-detection benchmark contains only positive bug instances. Consequently, accuracy on this benchmark is numerically equivalent to recall and does not measure false-positive behavior on truly behavior-preserving refactorings. We partially address this limitation through the complementary true-positive study in Section~\ref{sec:true-positives}, which evaluates 50 behavior-preserving refactorings. However, this study is smaller and does not fully characterize false-positive behavior across the broader space of valid refactorings, projects, and refactoring types. Thus, our results provide stronger evidence about detecting known refactoring bugs than about deployment-level false-positive rates.

\revisionTwo{
Another construct-validity threat concerns decoding configuration. Our additional \gptfive{} experiment with deterministic decoding (\(T=0\)) produced substantially lower behavioral-change accuracy than the default configuration, while compilation-error performance remained similar. Therefore, reported BC performance should be interpreted as conditional on the adopted inference settings rather than as an intrinsic property of the model itself. Different decoding configurations may lead to different accuracy--stability trade-offs.

The metamorphic-testing analysis also has limited power as evidence against memorization. The operators used in this study introduce behavior-preserving but mostly inert changes, such as comments, unused imports, unused fields, unused local variables, and auxiliary classes. These transformations are useful for evaluating robustness under controlled perturbations, but they do not prevent recognition by a model that may have seen the original code or structurally similar examples during training. Thus, our metamorphic results should not be interpreted as strong evidence against data contamination or higher-order clone recognition.
}

\subsubsection{External Validity}
%Our findings may not generalize beyond the studied setting. The benchmark focuses on Java refactoring bugs from popular IDEs and does not exhaustively cover all Java features, libraries, build systems, reserved keywords, refactoring tools, JDK versions, or ecosystems. Some features remain untested or underrepresented, such as \texttt{assert}, \texttt{volatile}, \texttt{transient}, \texttt{native}, and \texttt{strictfp}. The dataset is also imbalanced, since compilation errors outnumber behavioral changes. We therefore report per-category BC and CE metrics, but broader validation on larger, stratified corpora across languages, projects, refactoring kinds, IDEs, and JDK versions is needed.

The large-project diff-only study introduces additional external-validity threats. Since the model receives unified diffs rather than full project context, it may lack information about unseen call sites, imports, subclasses, dependencies, build files, framework callbacks, or external clients. For this reason, the prompt includes an \texttt{UNKNOWN} verdict, and we treat this experiment as a feasibility study rather than as a definitive benchmark. Human review reduces the risk of blindly trusting model outputs, but does not eliminate author bias in borderline cases involving API compatibility or behavioral equivalence.

\revision{Our findings may not generalize beyond the studied setting. The benchmark focuses on Java refactoring bugs from popular IDEs and does not exhaustively cover all Java features, libraries, build systems, reserved keywords, refactoring tools, JDK versions, or ecosystems. \revisionTwo{In particular, the results should be interpreted as evidence for Java refactoring correctness assessment only, and should not be generalized to other programming languages without further evaluation, since refactoring semantics, type systems, build infrastructures, and testing practices differ across ecosystems.} Some features remain untested or underrepresented, such as \texttt{assert}, \texttt{volatile}, \texttt{transient}, \texttt{native}, and \texttt{strictfp}. The dataset is also imbalanced, since compilation errors outnumber behavioral changes. We therefore report per-category BC and CE metrics, but broader validation on larger, stratified corpora across languages, projects, refactoring kinds, IDEs, and JDK versions is needed.}

\subsubsection{Conclusion Validity}
The statistical and comparative analyses should be interpreted cautiously. Statistical tests and confidence intervals are conditional on this dataset and model snapshot, and limited repeated runs for some proprietary models constrain the ability to estimate variance. Differences between models may partly reflect training data, inference infrastructure, hidden system prompts, model revisions, or reasoning style rather than inherent capability gaps. We mitigated this threat by evaluating all models on the same benchmark and by using paired statistical tests for the main model comparisons, but the conclusions remain conditional on the evaluated setting.
\revisionTwo{
In addition, the behavioral-change subset contains only \bugsBC{} instances distributed across \refactTypes{} refactoring types. Consequently, several refactoring types are represented by very few BC instances or none at all. Therefore, the BC results reported per refactoring type (Figure~\ref{fig:heatmap-refactoring-types}) should be interpreted as descriptive observations rather than statistically supported estimates.
}

\revisionTwo{
The large-project feasibility study also supports only limited conclusions. In this setting, 18 out of 44 cases were classified as \texttt{UNKNOWN}, showing that diff-only analysis often lacks enough context for definitive judgments. Moreover, the \texttt{YES} and disputed \texttt{NO - BEHAVIOR CHANGE} cases were assessed through author inspection of visible diff evidence rather than a mechanically established whole-project oracle. Therefore, this study should be interpreted as evidence about the feasibility and limitations of diff-only model-assisted triage, not as an accuracy benchmark for large-project refactoring correctness.
}

The mixture-of-experts, execution-time, and cost analyses also have limitations. The MoE analysis uses an optimistic OR-based criterion, measuring potential cross-model complementarity rather than a deployable decision procedure. Runtime and cost depend on hardware, provider, API latency, network conditions, batching, implementation overhead, and changing prices. These values are useful for understanding approximate trade-offs under our experimental conditions, but they should not be read as stable or universal benchmarks. Future work should increase sample sizes, expand repetitions, evaluate additional models, and investigate cost-aware and confidence-aware routing strategies.

Taken together, these threats delimit the strength of the empirical evidence. The study supports claims about observed bug-detection performance on known Java refactoring bugs, robustness under the tested metamorphic transformations, executable oracle generation for the evaluated benchmark, and complementary evidence from a limited set of behavior-preserving refactorings. It does not establish complete semantic reliability, general-purpose behavioral-equivalence checking, or deployment-ready correctness guarantees.
}

\section{Related Work}
\label{sec:relwork}

Opdyke and Johnson~\cite{Opdyke-SOOPPA-1990,Opdyke-PHD-1992} introduced the foundational concept of refactoring, while Roberts~\cite{Roberts-PHD-1999} pioneered the automation of basic refactoring operations. Subsequent work by Tokuda and Batory~\cite{Tokuda-ASE-2001} demonstrated that Opdyke's preconditions alone were insufficient to \revision{fully ensure} behavior preservation, highlighting the complexity of ensuring correctness in refactoring transformations. AlOmar et al.~\cite{DBLP:journals/infsof/AlOmarMNO21} performed a comprehensive systematic mapping study on behavior preservation during software refactoring, providing valuable insights into current practices, challenges, and research gaps in the field. Sch\"{a}fer et al.~\cite{Schafer-PLPV-2009} further emphasized the challenges of \revision{ensuring} correctness across all language constructs.
\revision{
Unlike these foundational works, which focus on formalizing, implementing, or proving refactorings, our work investigates whether foundation models can serve as oracle-like aids for detecting errors in existing refactoring transformations. We therefore position foundation models as a complementary correctness layer rather than as a replacement for traditional refactoring implementations or formal analyses.
}

\subsection{Testing Refactoring Implementations}

Daniel et al.~\cite{test-tools-fse07} proposed automated testing methodologies for Java refactoring implementations by generating programs and employing programmatic oracles to uncover faults such as overly weak preconditions. Their approach primarily detected issues manifesting as compilation errors through systematic program generation. Soares et al.~\cite{Soares-TSE-2013} later developed a systematic approach using \jdolly{} to generate Java programs and \saferefactor{}~\cite{saferefactor-ieee} to detect behavioral changes through automatically generated tests. Their methodology successfully identified 106 bugs, including compilation errors and behavioral changes, across 39 refactoring implementations. Drienyovszky et al.~\cite{quickcheck-erlang} similarly validated Erlang refactoring tools using automated property-based random testing based on formal refactoring specifications.
\revision{
Our work differs from these approaches in both objective and oracle design. Rather than testing refactoring implementations by generating new input programs or relying on hand-crafted properties, we evaluate whether foundation models can judge the correctness of already observed refactoring transformations. This allows us to analyze real bug reports from mature IDEs using zero-shot prompting, without relying directly on specialized generators, formal refactoring specifications, or executable test-generation infrastructure. These directions are complementary: dynamic and property-based oracles provide strong evidence when executable tests or formal properties are available, whereas model-based assessment may be useful for historical bugs, heterogeneous transformations, and scenarios where executable oracles are difficult to configure.
}

Dong et al.~\cite{dong-icse-2025} presented a ChatGPT-driven framework for testing refactoring implementations. Their approach synthesizes test programs using LLMs by mining feature libraries from bug reports and existing test cases, encoding preconditions, and guiding generation through prompt templates. The generated programs are then used for differential testing across multiple refactoring implementations, leading to the discovery of 115 Java refactoring bugs.
\revision{
In contrast, our goal is not to synthesize programs for testing refactoring implementations, but to evaluate whether a given transformation is correct. Our oracle directly classifies refactoring outcomes as behavior-preserving, behavior-changing, or compilation-invalid. These approaches are complementary: LLM-generated test inputs can expose new refactoring-implementation defects, while our model-based oracle can help assess whether the resulting transformations appear to preserve behavior.
}

Wang et al.~\cite{wang2024empiricalstudyrefactoringengine} conducted an extensive manual study analyzing 518 refactoring bugs across three widely used IDEs (\eclipse{}, \netbeans{}, \intellij{}), systematically identifying root causes, bug symptoms, and input-program characteristics. Their transferability analysis uncovered 130 previously unreported bugs, demonstrating the persistent nature of refactoring-related defects.
\revision{
Building on their dataset, we focus on two critical bug classes: compilation errors and behavioral changes. We refine the original artifacts by reconstructing executable Java input--output program pairs, compiling them to identify refactoring-induced compilation errors, and providing a JUnit test for each behavioral-change case to expose the behavioral difference. Our study then asks whether foundation models can detect these failures directly, providing an orthogonal perspective on refactoring-implementation reliability.
}

\subsection{Tools}

Sch\"afer and de Moor~\cite{Schafer-OOPSLA-2010} \revision{improved} Java refactoring support in \eclipse{} by formalizing transformations and implementing them in the JRRT tool, advancing soundness guarantees but requiring substantial language-specific engineering effort. Kim et al.~\cite{Kim-TSE-2014} showed that developers often apply refactorings manually, with the notable exception of Rename refactoring, emphasizing the need for more trustworthy automated support. Murphy-Hill and Black~\cite{Murphy-Hill-icse-2008} and Eilertsen and Murphy~\cite{refactoring-usability} further showed that refactoring-tool usability depends not only on correctness, but also on informative feedback, control, and developer trust.
\revision{
Our work complements these tool-oriented studies by focusing on an explanation-oriented validation layer for refactoring correctness. Rather than implementing refactorings or replacing existing IDEs, foundation models may help identify suspicious transformations and explain potential correctness risks. This aligns with prior calls for more informative refactoring feedback and suggests a possible role for foundation models in AI-assisted IDEs.
}

Horikawa et al.~\cite{horikawa2025agenticrefactoringempiricalstudy} presented a large-scale empirical study on refactoring activities performed by AI coding agents in real-world open-source Java projects. They find that agentic commits frequently include refactorings, mostly low-level and consistency-oriented edits.
\revision{
Our work complements their study by examining how foundation models can evaluate the correctness of refactorings, including those produced by AI agents. In this sense, our oracle-like analysis may serve as one validation component for agentic refactoring workflows.
}

\subsection{LLM-Assisted Refactoring}

Xu et al.~\cite{DBLP:conf/pldi/Xu0NH22} evaluated LLMs for code-related tasks, while Hou et al.~\cite{se-llms-2023} surveyed the broader use of LLMs in software engineering. Fan et al.~\cite{DBLP:conf/fose-ws/FanGHLSYZ23} highlighted open research problems at the intersection of LLMs and refactoring.
\revision{
More directly related to our topic, Martinez et al.~\cite{martinez2026llmrefactoringslr} conducted a systematic literature review of LLM-based refactoring research. Their review shows that the area is growing rapidly, but remains fragmented in terms of datasets, prompting strategies, tooling, and, especially, how correctness and accuracy are defined and measured. They also identify recurring challenges such as hallucinations, erroneous code generation, context loss, difficulties with complex refactorings, and scalability limitations.
Our study addresses one of these central gaps: systematic correctness assessment for LLM-assisted and tool-produced refactorings. Rather than asking models to generate refactorings, recommend opportunities, or improve code quality, we evaluate whether foundation models can serve as oracle-like components for already-applied transformations, detecting both compilation errors and behavioral changes.
}

White et al.~\cite{white2023chatgptpromptpatternsimproving} proposed prompt patterns for applying refactorings with ChatGPT, and AlOmar et al.~\cite{DBLP:conf/msr/AlOmarVMNO24} analyzed developer--AI conversations involving refactoring requests. Liu et al.~\cite{liu2024empiricalstudypotentialllms} studied GPT-4 and Gemini for identifying refactoring opportunities and recommending edits, introducing RefactoringMirror to reapply inferred transformations through a vetted refactoring implementation. Depalma et al.~\cite{DBLP:journals/eswa/DepalmaMHMA24} evaluated ChatGPT on Java refactoring tasks and reported both useful improvements and variability across identical prompts. Pomian et al.~\cite{DBLP:conf/sigsoft/PomianBDKBSBD24} proposed EM-Assist, an \intellij{} plugin that uses LLMs to suggest and rank \textit{Extract Method} refactorings. Other work has explored smell removal~\cite{cordeiro2024empiricalstudycoderefactoring}, Python simplification and transformation~\cite{DBLP:conf/apsec/ShirafujiOSMW23,DBLP:journals/pacmse/DilharaBBD24}, refactoring recommendation~\cite{ZHANG2025121753,DBLP:conf/issta/LiuWWXWLJ23,sbes-2025,sbes-2026}, and expert-guided or agentic refactoring~\cite{piao2025refactoringllmsbridginghuman,oueslati2026refagentmultiagentllmbasedframework}.
\revision{
These studies primarily focus on generating, recommending, applying, or improving refactorings. Our work is complementary: we focus on validating the correctness of transformations after they have been produced, including transformations that originate from an IDE, a developer, or an AI coding agent. This makes our oracle-style approach a potential downstream validation layer for LLM-assisted refactoring pipelines.
}

\revision{
\subsection{Comparison with Traditional Baselines}
\label{sec:comparison-traditional}

To position foundation models with respect to executable approaches, we compared our method against two traditional baselines on the same benchmark: Java compilation checking for compilation-error detection and \saferefactor{}~\cite{saferefactor-ieee} for behavior-preservation assessment.
For compilation errors, \openjdk{} correctly identifies all CE cases in our dataset, as expected, since compilation validity is ultimately determined by the Java language implementation and compiler version. \saferefactor{} performs both compilation checking and dynamic behavior-preservation assessment through affected-entity analysis and automatically generated tests. In our setting, \saferefactor{} correctly handles 221 out of \bugs{} benchmark cases, corresponding to 97.8\% accuracy.
The five incorrect cases are IDs~122, 137, 140, 189, and~190. In our benchmark, these are compilation-error cases: the source program compiles, but the resulting program does not. However, \saferefactor{} reported that both the source and resulting programs failed to compile. As a result, it misclassified the failure as an invalid source-program setup rather than as a compilation error introduced by the refactoring. Thus, these failures indicate that \saferefactor{} could not properly analyze the transformation under its required execution environment.
\saferefactor{} depends on \randoop{}~\cite{randoop-icse07} for test generation and requires a Java~8 execution environment. However, some benchmark programs use newer Java features, such as \texttt{var}, switch expressions with \texttt{yield}, or static nested classes in contexts supported only by later Java versions. Consequently, these failures are caused by version incompatibilities in the execution environment rather than by the refactoring transformations themselves.

In comparison, \gemini{} achieved a correct classification rate of \geminiAcc{}, slightly outperforming \saferefactor{} on this benchmark, while \gptfive{} and \gptoss{} detected \gptfiveAcc{} and \gpttossAcc{} of the bugs, respectively, in the first-run setting. Thus, although \saferefactor{} remains highly competitive and provides executable evidence when its Java~8-based toolchain is applicable, the strongest foundation model was able to match or exceed its observed accuracy while also handling newer Java features present in our dataset.
This comparison highlights a key trade-off. Compilers and tools such as \saferefactor{} provide strong executable evidence and are generally deterministic, fast, and inexpensive when applicable. Foundation models, in contrast, are less deterministic and do not provide formal guarantees, but they may be more adaptable with respect to language evolution, can produce natural-language explanations, and can be applied in settings where dynamic test generation or project configuration is difficult. In particular, \saferefactor{} achieved 100\% accuracy on the applicable BC cases, so the main advantage of foundation models in this comparison lies not in higher BC accuracy, but in their ability to handle newer Java features, provide natural-language explanations, and operate without requiring a \saferefactor{}-compatible test-generation environment. We therefore view foundation models as complementary oracles rather than replacements for traditional analysis-based approaches.
}

\revisionTwo{
JRRT is highly relevant to refactoring correctness research, but it is not a directly comparable baseline for our evaluation. JRRT is a refactoring engine that implements specific refactoring transformations and attempts to prevent incorrect transformations through refactoring-specific preconditions~\cite{Schafer-OOPSLA-2010}. In contrast, our approach receives the original and refactored programs as input and assesses whether the resulting transformation preserves behavior or introduces a compilation error, regardless of how the transformation was produced. Therefore, we consider \saferefactor{} the more appropriate traditional baseline because it addresses the same validation problem: post-refactoring validation over a pair of program versions. Prior work used \saferefactor{} to uncover compilation errors and behavioral changes in JRRT implementations~\cite{Soares-TSE-2013}, illustrating that validation tools and refactoring engines play complementary roles. Similarly, our approach is intended to validate the outputs of refactoring engines such as JRRT, rather than to replace them. Table~\ref{tab:traditional-comparison} summarizes the comparison between the evaluated foundation models and the traditional approaches discussed in this section.
}

\begin{table}[t]
\centering
\caption{\revisionTwo{Comparison between traditional post-refactoring validation approaches and foundation-model oracles.}}
\label{tab:traditional-comparison}
\small
\begin{tabular}{lrr}
\hline
\textbf{Approach} & \textbf{CE Acc.} & \textbf{BC Acc.} \\
\hline
\openjdk{}           & 100.0\% & --- \\
\saferefactor{}      & 97.3\%  & 100.0\% \\
\gptoss{}            & 77.8\%  & 92.7\% \\
\gemmaNew{}             & 96.2\%  & 97.6\% \\
\gptfive{}           & 94.1\%  & 92.7\% \\
\gemini{}            & 100.0\% & 97.6\% \\
\claude{}            & 95.1\%  & 92.7\% \\
\hline
\end{tabular}
\end{table}

\subsection{Robustness}

Sallou et al.~\cite{sallou2024breaking} examined systemic risks associated with deploying LLMs in software engineering contexts, including data leakage, reproducibility challenges, and dependence on closed-source models. They propose mitigation guidelines emphasizing rigorous empirical validation. Zhang et al.~\cite{alucinacao} surveyed LLM hallucinations and classify them into input-, context-, and fact-conflicting categories.
\revision{
Our study explicitly considers these risks. We evaluate repeated attempts, temperature sensitivity, and metamorphic transformations to assess stability and sensitivity to input-preserving variations. We also analyze malformed outputs, incorrect semantic verdicts, and cases in which generated tests fail to compile or fail to expose behavioral differences. These analyses do not eliminate threats such as data leakage, hallucination, or provider-specific behavior, but they provide evidence, within our setting, about when model predictions are stable and when they remain fragile.
}

\section{Conclusions}
\label{sec:conclusion}

In this article, we investigated the \revision{potential} of foundation models to \revision{serve as oracle-like components} for detecting refactoring bugs, including both behavioral changes and compilation errors introduced by program refactorings. Our empirical results show that these models can provide informative signals about refactoring correctness across real-world bugs reported over more than a decade in widely used Java IDEs (\intellij{}, \eclipse{}, and \netbeans{}). Using zero-shot prompting without task-specific training, the models were evaluated across multiple refactoring types and Java language features, while also producing natural-language rationales that may help developers inspect their predictions.
\revision{
The evaluation reveals substantial performance differences between models. In the first-run setting, \gptoss{} achieved \gpttossAcc{} overall accuracy, while \gptfive{}, \claude{}, and \gemini{} reached $93.8\%$, $94.7\%$, and $99.6\%$, respectively. Among the evaluated \revisionTwo{open-weight} models, \gemmaNew{} achieved the strongest result, correctly classifying $96.5\%$ of the cases. Repeated sampling further improved cumulative coverage for \gptoss{} and \gptfive{}, although majority voting was less effective than using multiple attempts to recover at least one correct answer.
}

\revision{
The execution-time and cost analyses highlight deployment-relevant trade-offs. Smaller open-weight models were faster but less accurate, while stronger open-weight models such as \gemmaNew{} achieved high accuracy at higher runtime cost. Thus, lightweight or interactive triage is practical only for some of the evaluated open-weight models, as larger models such as \gemmaNew{} and \qwen{} incurred substantially higher runtimes in our environment. \gemini{} and \gptfive{} offered the best latency--accuracy balance among the strongest models in our setting. Local cloud execution of \gptoss{} may be useful for exploratory analyses or first-pass triage, whereas stronger API-based models may be preferable when higher accuracy, stability, or low latency is required. These estimates are indicative rather than definitive, since prices and inference infrastructure are likely to evolve.
}

Our findings also indicate that LLMs \revision{may} help analyze refactoring transformations before they are applied to an entire codebase. This preventive use case may be useful in large projects, where full builds or test suites may be costly, slow, or dependent on complex environments. The large-project feasibility study reinforces this potential, but also shows that diff-only reasoning often requires an \texttt{UNKNOWN} outcome because decisive evidence may lie outside the visible patch.
An additional insight concerns the impact of language evolution on refactoring-validation techniques. Traditional tools that rely on automatic test generation, such as \saferefactor{}, require continual engineering effort to support new language constructs and are constrained by the test generators on which they depend. In contrast, foundation models were able, \revision{in our dataset}, to analyze refactorings involving newer Java features and provide explanations tied to their semantics. This suggests that LLM-based oracles may offer a flexible complement to traditional validation tools as programming languages evolve, although this flexibility does not provide formal guarantees and must be validated per setting.
\revision{
Overall, we do not view foundation models as replacements for existing validation tools, but rather as complementary mechanisms. Given \saferefactor{}'s 100\% accuracy on the applicable BC cases, the main advantages of foundation models in this comparison are broader applicability to newer Java features and natural-language explanation support, reinforcing their complementary rather than superior role. Deterministic techniques should be preferred whenever a property can be checked directly, especially for compilation errors. At the same time, foundation models provide a unified analysis interface and useful signals in settings where traditional tools are incomplete, costly, or difficult to configure. Hybrid workflows combining deterministic tooling with LLM-based analysis and explanations appear to be a promising direction for practical adoption.
}

Nevertheless, several limitations remain. The models are sensitive to prompt design, context representation, decoding configuration, and the adopted definition of behavioral equivalence. Context-window limitations restrict direct analysis of large projects, and model non-determinism introduces stability concerns for production settings. Human oversight remains crucial, especially when model explanations are plausible but insufficiently grounded. \revisionTwo{
Moreover, our evaluation focuses on Java refactoring tasks and primarily on general-purpose foundation models. Therefore, the reported results should not be interpreted as evidence that the observed performance generalizes to other programming languages or to code-specialized foundation models. This also means that our results do not isolate whether performance is primarily driven by general reasoning capabilities, broad code exposure during pretraining, or software-engineering-specific model specialization.
}

For future work, we plan to expand the evaluation to additional programming languages, larger and more diverse codebases, and a broader set of refactoring scenarios. \revisionTwo{We also plan to evaluate code-specialized foundation models to better understand the relative contributions of general reasoning and software-engineering-specific knowledge to refactoring correctness assessment.} \revision{We also intend to investigate tighter integration with compilers, static analyzers, test-generation tools, and retrieval-augmented project context, particularly to improve systematically hard compilation-error cases and ambiguous diff-only scenarios.} In addition, we plan to explore \revisionTwo{one-shot, few-shot, and chain-of-thought prompting,} structured prompting, \revision{temperature and decoding-parameter sensitivity for larger closed models such as \gptfive{},} ensemble, mixture-of-experts, and agentic strategies to improve stability and reduce attempt-sensitive errors. \revisionTwo{We also plan to extend the repeated-sampling protocol to all evaluated models to support more uniform stability comparisons across model families.} 
\revision{Finally, we see a promising opportunity in developing lightweight refactoring-aware checkers and integrating foundation models into AI-augmented IDEs for interactive refactoring assistance, validation, and decision support.}

\section*{Acknowledgments}
We thank the anonymous reviewers for their insightful suggestions. This work was partially supported by CNPq (306026/2026-0, 408040/2025-4, 403719/2024-0), CAPES (88887.313474/2026-00), FAPESQ-PB (268/2025), and INES.IA (National Institute of Science and Technology for Software Engineering Based on and for Artificial Intelligence), \url{http://www.ines.org.br}, CNPq grant 408817/2024-0. 

% \subsection*{Ethical Approval}
% Not applicable.

% \subsection*{Consent to Participate}
% Not applicable.

% \subsection*{Consent for Publication}
% Not applicable.

% \subsection*{Author Contributions}
% Rohit Gheyi: Methodology, Investigation, Conceptualization, Software, Data Curation, Writing -- Original Draft.

% \noindent Rian Melo: Software, Data Curation, Validation, Writing -- Review and Editing.

% \noindent Jonhnanthan Oliveira: Software, Data Curation, Validation, Writing -- Review and Editing.

% \noindent Márcio Ribeiro: Methodology, Investigation, Conceptualization, Writing -- Original Draft.

% \noindent Baldoino Fonseca: Methodology, Investigation, Conceptualization, Writing -- Original Draft.

% \subsection*{Data Availability}
% All data and artifacts supporting the findings of this study are publicly available on Zenodo: \url{https://zenodo.org/records/21910261}.

\end{document}